\documentclass[a4paper,12pt,rmp,superscriptaddress]{revtex4}
\usepackage{latexsym}
\usepackage{setspace}
\usepackage{color}
\usepackage{hyperref}
\usepackage{fancyhdr}
\usepackage{graphicx}
\usepackage{wrapfig}

\usepackage[margin=2cm]{geometry}

\begin{document}

\title{Physical descriptions of the bacterial nucleoid at large
  scales, and their biological implications.}

\author{Vincenzo~G. Benza}
\affiliation{Dipartimento di Fisica e Matematica, Universit\`a
  dell'Insubria, Como, Italy}
\author{Bruno Bassetti} \affiliation{Universit\`a degli Studi di
  Milano, Dip.  Fisica, Via Celoria 16, 20133 Milano, Italy}
\affiliation{I.N.F.N. Milano, Via Celoria 16, 20133 Milano, Italy}
\author{Kevin~D. Dorfman}
\affiliation{Department
  of Chemical Engineering and Materials Science, University of
  Minnesota - Twin Cities, 421 Washington Ave. SE, Minneapolis,
  Minnesota 55455, United States}
%
\author{Vittore~F. Scolari}
\affiliation{Genomic Physics
  Group, UMR 7238 CNRS ``Microorganism Genomics''}
\affiliation{University Pierre et Marie Curie, 15 rue de l'\'{E}cole
  de M\'{e}decine Paris, France}
\author{Krystyna Bromek}
  \affiliation{Cavendish Laboratory and
  Nanoscience Centre, University of Cambridge, Cambridge CB3 0HE,
  U.~K.}
\author{Pietro Cicuta}
  \affiliation{Cavendish Laboratory and
  Nanoscience Centre, University of Cambridge, Cambridge CB3 0HE,
  U.~K.}
\author{Marco Cosentino Lagomarsino}
\affiliation{Genomic Physics
  Group, UMR 7238 CNRS ``Microorganism Genomics''}
\affiliation{University Pierre et Marie Curie, 15 rue de l'\'{E}cole
  de M\'{e}decine Paris, France}

\begin{abstract}
  Recent experimental and theoretical approaches have attempted to
  quantify the physical organization (compaction and geometry) of the
  bacterial chromosome with its complement of proteins (the
  nucleoid). The genomic DNA exists in a complex and dynamic
  protein-rich state, which is highly organised at various length
  scales. This has implications for modulating (when not directly
  enabling) the core biological processes of replication,
  transcription, segregation.  We overview the progress in this area,
  driven in the last few years by new scientific ideas and new
  interdisciplinary experimental techniques, ranging from high space-
  and time-resolution microscopy to high-throughput genomics employing
  sequencing to map different aspects of the nucleoid-related
  interactome.  The aim of this review is to present the wide spectrum
  of experimental and theoretical findings coherently, from a physics
  viewpoint.
  In particular, we highlight the role that statistical and soft
    condensed matter physics play in describing this system of
    fundamental biological importance, specifically reviewing classic
    and more modern tools from the theory of polymers.
    We also discuss some attempts towards unifying interpretations of
    the current results, pointing to possible directions for future
    investigation.
\end{abstract}

\maketitle

\setstretch{1}


\section{Introduction}

The demarcation line between a ``physical'' and a ``biological''
system is rapidly becoming anachronistic, to the point that it
possibly hinders research in both disciplines. This is particularly
true in the context of gene expression, which is a fundamental process
in biology at the molecular level, common to all life on earth: the
genetic code is read out (``transcribed'') from DNA and written into
RNA, which, in the case of messenger RNA, is then translated into
proteins. In order for the right number of proteins to be produced in
response to changes in environment, internal states, and stimuli, all
cells are capable of tightly regulating this sequence of
events~\cite{Alberts2008_book}.  The physico-chemical implementation
of this gene regulation process takes place primarily through specific
DNA-binding interactions of transcription factors, which repress or
promote transcription.

It is becoming very clear that the genome's conformational properties as
a polymer come into play in the processes involved in the regulatory
fine-tuning of gene expression, in particular its topological,
chemical, geometric and mechanical properties. These properties can
influence the activity of the transcription factors and can play a
role in the coordination of a large scale cellular response. Evidence
for this level of regulation has been found in the different kingdoms
of life.  We focus here on the efforts to describe the genome's
physical state in the case of bacteria, where it is (perhaps
surprisingly) less explored than for higher life forms.

Understanding this problem requires stringent, quantitative
experiments with the standards of physics, together with up-to-date
physical models and arguments.
Since considerable knowledge has been developed in polymer science
over the last 50 years, it is very tempting to try to apply this
knowledge to understand the energy and time scales involved in
maintaining the DNA at the same time compact and accessible inside a
cell, and the role of its geometry, structure and compaction in gene
regulation.

%
%
%

This review discusses challenges that arise in the biological arena,
 in which mature experimental and theoretical tools from
the physical sciences might now allow significant progress. The review
is primarily aimed at our colleagues in the physical sciences: it
should communicate a feel for the main questions and the main
challenges, and our understanding that a comprehensive physical
approach is possible and necessary at this point. We will not explain
the physical models in great technical detail, and we hope this work
will also be of interest to biologists who could take the references
given here as a starting point and a ``compass'' in order to evaluate
different modelling approaches. Ultimately, we believe that optimal
progress in this area will take place in collaboration, and this
review might contribute to establishing a common ground and language.

\begin{center}
  \line(1,0){250}
\end{center}

\vspace{0.5cm}
\begin{center}
  \textbf{BOX~1: Main factors affecting nucleoid organization.}
\end{center}
\vspace{0.5cm}
\label{pageboxMAINFACTORS}

\textbf{Supercoiling by topoisomerases.} These enzymes affect the
winding of DNA, and play a role in the creation of the observed
branched structure of plectonemic loops along the genome.
\textbf{Nucleoid associated proteins (NAPs).} These proteins bind to
DNA with different specific modes, each responsible for a different
aspect of organization, ranging from double-strand bridging to
nucleoprotein filament formation.
\textbf{Confinement.} Millimeters of genomic DNA are confined within
the small cell volume of a bacterium.
\textbf{Molecular crowding.} There is a high concentration of macromolecules
present in the cytoplasm, This factor could affect nucleoid
organization at different levels, for example creating a general effective
self-attraction favoring collapse, and strong depletion attraction
between large objects, such as ribosomes.
\textbf{Replication and transcription.} These are the non-equilibrium
processes of DNA and mRNA production that continuously take place in
dividing cells. The first affects the sheer amount of genome present
and supercoiling and makes both highly nonsteady, the second can
create uneven ribosome concentrations. Both contribute to
non-thermal force and displacement fluctuations of the chromosome.

\begin{center}
\line(1,0){250}
\end{center}


\section{Background}

Bacteria are single cell organisms of fundamental importance in
 nature and to mankind. They are ``prokaryotes'', in the
etymological sense that they lack a nucleus as a compartment
 enclosed by a membrane (in contrast to the
``eukaryotic'' cells that make up animals and plants).  Indeed, all the
DNA, RNA, and proteins in the bacterial cell are always present
together in the same single compartment.
While the classical picture of bacteria (still implicitly adopted by
many theoretical and experimental investigators) views them as little
more than a bag (or ``well-stirred reactor'') of proteins and DNA, it
is now clear that this tenet is flawed on many different
levels. Despite their lack of membrane-bound organelles, bacteria have
a high degree of intracellular spatial organization, related to most
cellular processes.
Perhaps the highest organized structure of the bacterial cell is the
genome itself.

In most bacteria, the chromosome is a single circular DNA molecule
confined by the cell membrane. In \emph{E.~coli} (the best studied
bacterial system), the chromosome consists of about $4.7$ million base
pairs (bp) and has a total length of $1.5$
mm~\cite{Trun1998,Stavans2006}. Bacterial DNA is organized into a
specific structure called the nucleoid, which is composed of DNA, RNA
and proteins, and occupies a well-defined region of the
cell~\cite{Sherratt2003}.
The nucleoid is organized by a set of nucleoid associated proteins, or
``NAPs'' (such as Dps and transcription factors Fis, H-NS, IHF, HU), which
can modify the shape of the DNA both at local and global
levels~\cite{LNW+06,Ohniwa2011}.
Since the linear size of the genomic DNA is orders of magnitude larger
than the length of the cell, it must be packaged and organized in such
a way that the resulting structure is compact, while still allowing
the primary information processing, genome replication and gene
expression~\cite{Thanbichler2005}. In fact, recent evidence strongly
suggests that changes in chromosome architecture can directly affect
the accessibility and activity of the regulatory proteins at the local
level as well as at larger scales. In addition, the genome can
efficiently control gene expression by changing the way DNA-binding
regulatory proteins can access their target sites via the chromosome
architecture~\cite{Dillon2010}.
Historically, the nucleoid has been visualized by transmission
electron microscopy (TEM), phase-contrast microscopy and confocal
scanning light microscopy; an overview of early microscopic
visualization of the nucleoid can be found
in~\cite{Robinow1994}. These studies showed that the nucleoid mostly
occupies a separate subcellular region, without being bound by a
membrane, and that thin DNA threads extrude from this region.

The double-stranded genomic DNA is generally torsionally constrained
in bacteria, typically in such a way that its linking number (the
number of times each single strand of DNA winds around the other) is
lower than in the relaxed configuration.
In biological words this is referred to as ``negatively supercoiled'',
and the specific difference in linking number is called superhelical
density.  The name is due to the fact that supercoiled DNA develops
nonzero writhe, or ``supercoils'', \emph{i.e.} it is wrapped around
itself in the manner of a twisted telephone cord.
This torsional constraint has important consequences for gene
expression, as the mechanical stress carried by a negatively
supercoiled configuration can locally weaken the interaction between
the two strands. The resultant breaking of base-pair bonds between the
two helices is required for the initial steps of transcription, as
well as DNA replication and recombination. Since most DNA-binding
proteins bind to DNA sensitively to the arrangement of the two
strands, and in particular to their average distance, negative
supercoiling usually also facilitates protein
binding~\cite{Dillon2010}.
The level of supercoiling is tightly regulated by the cell, and it can
be changed by the action of specific enzymes such as topoisomerases
and gyrases.

The average supercoiling is generally negative in bacterial cells (except
for thermophilic bacteria~\cite{Confalonieri1993}).
In the case of \emph{E. coli}, the supercoiling superhelical density
is maintained around the value $\sigma = -0.025$, where the supercoil
density $\sigma$ is defined as the relative change in linking number
due to the winding (or unwinding) of the double helix, where negative
values of $\sigma$ correspond to an unwound helix~\cite{Stavans2006}.
This value is set by the constraining action of DNA-binding proteins
and the combined activity of topoisomerases. Deviations larger than
$20\%$ to either side of $\sigma$ are detrimental to cell growth. For
example, overwinding causes the formation of DNA structures that
impede transcription and replication, while excessive underwinding
leads to poor chromosome segregation~\cite{Stavans2006}.


The most important \emph{features linking nucleoid organization and
  cell physiology} are summarized here, and discussed further in the
review. See also BOX~1 on page \pageref{pageboxMAINFACTORS} for a
  brief description of the main factors affecting nucleoid
  organization.
\begin{itemize}
\item[(i)] At large scales, it is seen from \emph{ in vitro}
  experiments that the nucleoid is composed of topologically unlinked
  dynamic domain structures; these are due to supercoiling forming
  plectonemes and toroids~\cite{marko}, and stabilized by nucleoid-associated
  proteins, for example Fis. This combination of effects gives the
  chromosome the shape of a branched tree-like polymer visible from
  TEM~\cite{Postow2004,Skoko2006,Kavenoff1976}. Topological domains
  are thought to be packaged during replication, otherwise it appears
  that their boundaries are ``fluid'' and randomly
  distributed~\cite{Postow2004,Thanbichler2006}.
\item[(ii)] Strong compaction is experimentally observed \emph{in
    vivo}, and possibly arises from confinement within the cell
  boundaries, but also from various factors such as molecular
  crowding~\cite{Vries2010} and supercoiling~\cite{Stuger2002}. The
  degree of compaction changes with the cell's growth conditions
  and in response to specific kinds of stress.
  The \emph{E.~coli} chromosome, with a linear size of $1.5$mm,
  occupies a volume of 0.1-0.2 $\mu$m$^3$ (the bare DNA volume is about
  a factor 20-30 smaller), which brings the need to study the folding
  geometry of the nucleoid~\cite{Stavans2006};
\item[(iii)] Supercoiling and nucleoid organization play an important
  role in gene expression~\cite{Dillon2010}.  For example, during
  rapid growth, while several chromosome equivalents are present in
  the cell due to multiple ongoing replication cycles, an increased
  production of some nucleoid-associated proteins is observed,
  compacting the chromosome and probably giving rise to specific
  transcription patterns in a way that is not yet fully characterized.
\item[(iv)] RNA polymerase, the nucleoid-bound enzyme responsible for
  gene transcription, is concentrated into transcription foci or
  ``factories,''
  which can affect the nucleoid structure by bringing together distant
  loci~\cite{JC06,GHH+05}.
\item[(v)] For cells that are not replicating the genome, the position
  of genetic loci along the chromosome is linearly correlated with
  their position in the
  cell~\cite{Viollier2004a,Breier2004,Wiggins2010}. The exact
  subcellular positioning of different loci varies in different
  bacteria~\cite{Toro2010}.
\item[(vi)] During replication, daughter chromosomes demix and
  segregate before cell division. In \emph{ E.~coli}, the two arms of
  the chromosome are segregated in an organized
  $<$left-right-left-right$>$ asymmetric fashion from the center of
  the cell~\cite{WLP+06,NYH+07,Toro2010}. Replicated chromosomal loci
  are thought to be immediately recondensed, as they appear to
  preserve the linear arrangement while they are moved in opposite
  directions to assume their final position in the incipient daughter
  cell~\cite{Thanbichler2005,Thanbichler2006}.
\end{itemize}

In short, the nucleoid's physical organization plays a major role in the
most important cellular processes, such as cell division, DNA replication,
and gene expression.
Explaining these links is a long-standing open problem in
microbiology. Today, it can be revisited with new quantitative
experiments, including both the ``-omics'' approaches and more
sophisticated and controlled experimental techniques allowing the
analysis of nucleoids both \emph{in vitro} and \emph{in vivo}.
This is paralleled by a renewed interest in the quantitative
characterization of bacterial
physiology~\cite{Scott2011,Scott2010,Zaslaver2009}, of which the
nucleoid constitutes a fundamental part.
Also note that the relative chromosomal positioning and the orientation of
genes are subject to natural selection on evolutionary time-scales as
shown by the comparison of the genetic maps of
different bacteria~\cite{Ochman1994,Roc08}.

Consequently, the field is blossoming with a new wave of studies,
hypotheses and findings. While many new pieces of evidence are
available, the challenge of building coherent pictures for physical
nucleoid organization and its role in the different cell processes
remains open.
Our scope here is to present the main hypotheses and the experimental
and modeling tools that have been put forward in order to understand
the physical aspects of nucleoid organization, and give the interested
reader an ordered account of the known facts.

\section{Measurements}
\label{sec:meas}

We start by reviewing some of the salient experimental findings at the
scale of whole-nucleoids, with a particular emphasis on the more
recent results.  The evidence that we will discuss emerges from a
combination of experimental biology, biophysics, high throughput
biology and bioinformatic approaches (see BOX~2 on page \pageref{pagebox}).
Most of the studies reviewed here are based on \emph{E.~coli} or
\emph{Caulobacter}.

\newpage

\begin{center}
  \line(1,0){250}
\end{center}
\textbf{BOX~2: Several available experimental techniques can probe the
  nucleoid at large scales. }

\label{pagebox}

\begin{center}
\includegraphics[width=0.54\textwidth]{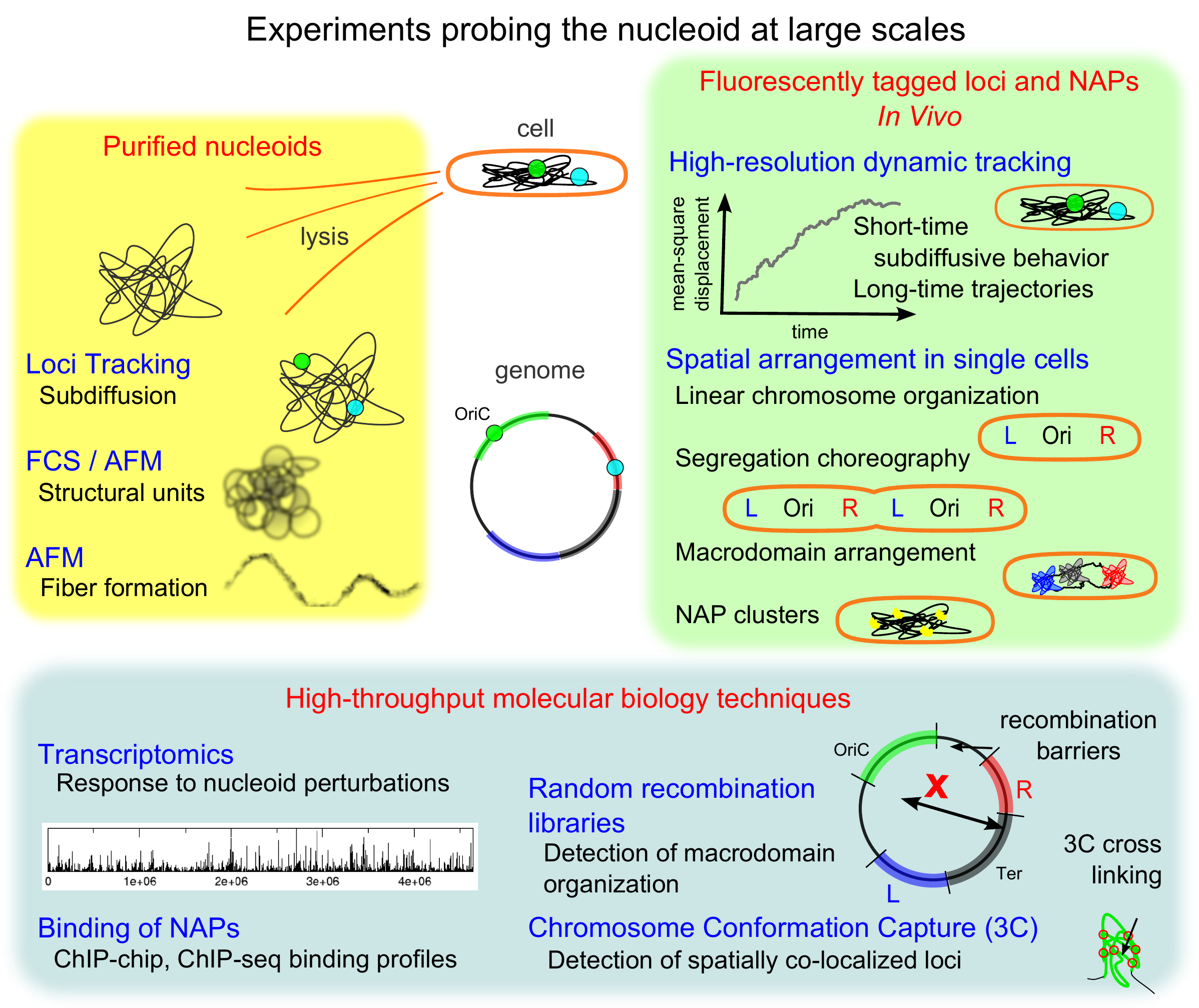}
\end{center}

Advanced microscopy together with cell-biological techniques yield
information about structure and dynamics of the nucleoid (Right).
{\bf High-resolution tracking} of tagged loci allows measurement of
the local viscoelastic properties of the nucleoid.  {\bf Static
  configurations} of chromosomal loci in fixed cells allow
determination of their spatial arrangement within nondividing cells
and following cell division.  Dynamic tracking at long time scales
gives information on the chromosome's {\bf ``choreography''}
followed  over a cell cycle
 and on the macrodomain subcellular arrangement.
Fusions of NAPs with fluorescent proteins also enable evaluation
of their {\bf localization} within the nucleoid.

Nucleoids can be purified and manipulated outside of a cell in order
to access more directly their biophysical properties (Left).  This
procedure implies release of confinement and crowding, and dilution of
binding proteins. As a result, purified nucleoids are several times
larger in radius than the size of a cell. Traditionally, purified
nucleoids were imaged by electron microscopy, showing a ramified
plectonemic structures. More recently, investigators have concentrated
on characterizing their organization as polymers {\bf tracking of
  labelled loci}, fluorescence correlation spectroscopy({\bf FCS}) and
probing them mechanically ({\bf AFM}).

In addition, information on nucleoid organization can be obtained
from high-throughput experimental datasets (Bottom).  {\bf
  Transcriptomics} (using sequencing or microarrays) can be used to
probe transcriptional response to nucleoid perturbations, such as NAP
deletions, changes in the average level of supercoiling, or local
release of a plectonemic loop.  Another important source of data is
{\bf protein occupancy}, for example by NAPs, and its correlation with
gene expression. This information is obtained both by microarray
(CHiP-chip) and by next-generation sequencing techniques
(ChIP-seq). {\bf Recombination} has been used to define macrodomains,
as compartments within which recombination between chromosomal
segments was more likely than recombination with segments laying
outside of the compartment. Finally, {\bf Chromosome Conformation
  Capture} (3C) techniques probe the spatial vicinity of pairs of
chromosomal loci in the (average) cell. They can be combined with
sequencing in order to produce high-throughput data sets (Hi-C).

\begin{center}
\line(1,0){250}
\end{center}

\newpage

\paragraph{Chromosome spatial arrangement and compartmentalization. }

Strikingly, the intracellular localization of a given chromosomal
locus in a cell is remarkably deterministic, as revealed by
fluorescent tagging of chromosomal loci on \emph{E.~coli} and
\emph{Caulobacter
  crescentus}~\cite{NYH00,Viollier2004a,WLP+06,Liu2010,NOY+06,Wiggins2010}.
The series of chromosome segments is localized along the long axis of
the cell in the same order as their positions along the chromosome
map, with the interlocus distance typically linearly proportional to
(arclength) genomic distance. The precise location of individual loci
varies in the known bacteria, and might depend on DNA-membrane tethering
interactions~\cite{Toro2010}.
A recent microscopy study on \emph{E.~coli}~\cite{Meile2011}
considered the positioning of the chromosome in the short-axis section
of the cell.  They found that the Ter region occupies the periphery of
the nucleoid, at a larger distance from the longitudinal axis with
respect to the rest of the chromosome.
In newborn or  non-replicating cells, the two chromosome arms are
spatially arranged such that loci on the left arm of the chromosome lie in one
half of the cell and loci on the right arm
lie in the opposite half, with the replication origin between them.
It is tempting to interpret the resulting sausage-shaped structure as
a chromosomal  ``fiber.''
In a recent study~\cite{Wiggins2010}, the cell-to-cell variability of
loci positioning in non-replicating cells was used to estimate an
internal elasticity, which, perhaps not surprisingly, appears to be
much higher than expected from a naive estimate for a linear polymer.

Even more recently~\cite{Umbarger2011}, high-throughput Chromosome
Conformation Capture (3C) techniques have been used in combination
with live-cell fluorescent tagging of loci, in order to determine the
global folding architecture of the \emph{Caulobacter crescentus}
swarmer cell genome.  These data indicate that a chromosomal fiber
exists also in this case, spanning the whole chromosomal
ring. Additionally to the linear spatial arrangement of loci according
to their chromosomal coordinate, loci of the left chromosomal arm tend
to be very proximal to symmetric loci on the right arm.
The resulting structure is a compressed ring-like fiber, which, the
authors argue, typically takes an eight shape, free to roll around the
long cell axis.
They also find that the symmetry in the cross-chromosomal arm
interactions is determined by the protein-dense attachment point to
the cell membrane at the old pole of the cell, triggered by the
binding of the ParB protein to its target parS binding sites.
Moving the parS pole-anchoring site by 400 Kb along the chromosome
(but not the replication origin) determines a sliding of the whole
interaction structure, as in a tank crawler.
This sliding is slightly asymmetric, suggesting the presence of
supplementary attachment points between chromosomal arms or between
the chromosome and the cell body.

\begin{figure}
  \centering
    \includegraphics[width=0.8\textwidth]{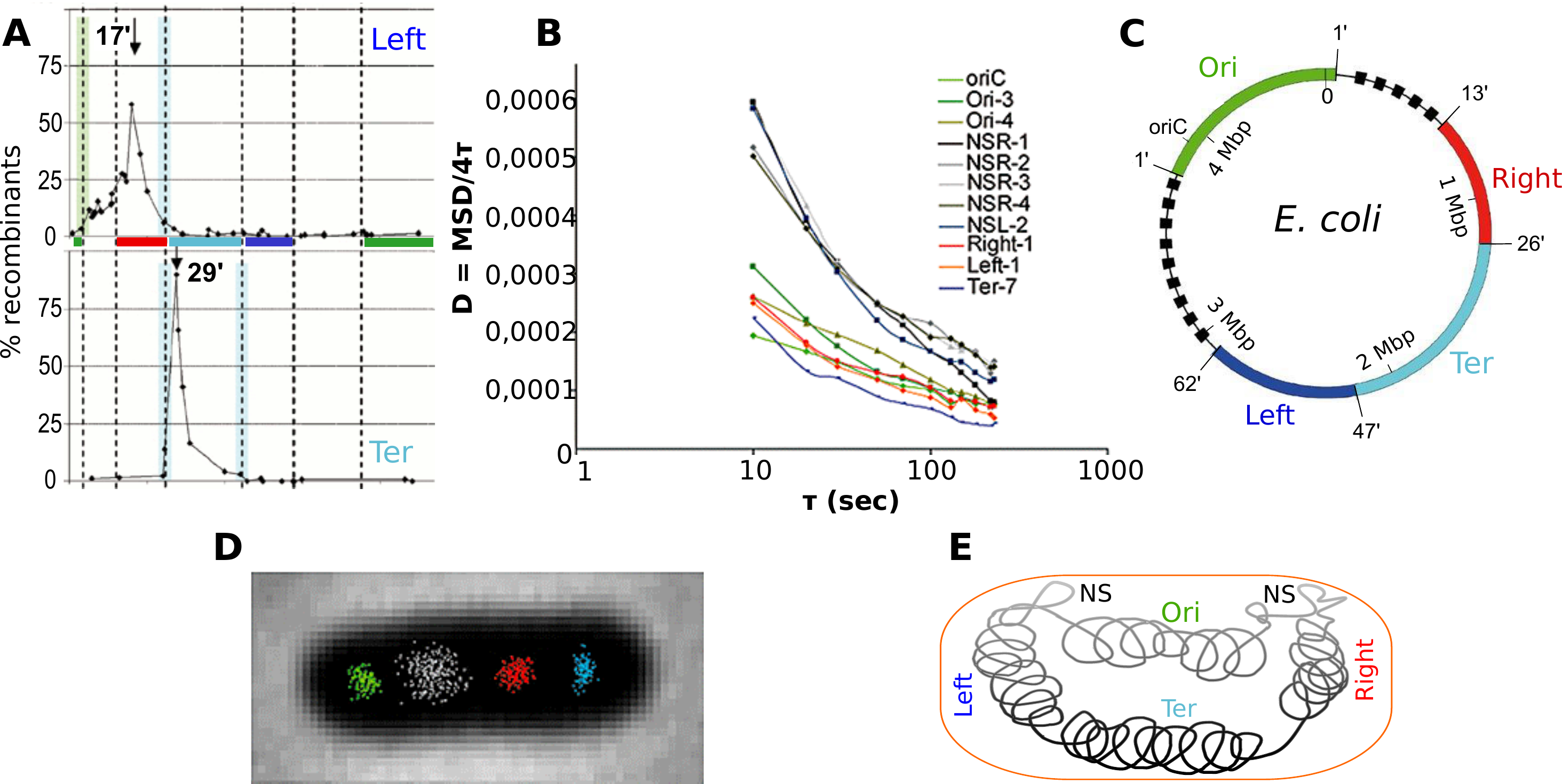}
    \caption{ \emph{Chromosome compartmentalization and spatial
        arrangement of genes.}
      (A) An example of the data that was used for the genetic
      definition of Right and Ter macrodomains, reprinted by
        permission from Macmillan Publishers
        Ltd. from~\citet{VPR+04}, $\copyright 2004$.
        The plot shows the frequency of recombination events between
        the genetic position indicated by the arrow and other
        positions probed along the genome (The x-axis indicates the
        base pair coordinates of the chromosome).  This should be
        uniform for a well-mixed polymer. On the contrary, experiments
        show a highly non-uniform pattern, compatible with a
        compartmentalized structure with clear boundaries.
 Dashed vertical lines indicate the delimitation of
macrodomains. Colored bars show the extent of the defined
macrodomains.
(B)
Effective diffusion constants obtained from loci tracking experiments,
 reprinted by permission from John Wiley and Sons
  Ltd. \citet{EMB08}, $\copyright 2008$ . The plot shows clear
differences between the
behaviour of loci placed within macrodomains and in non structured
regions (NSR1-4,NSL-2).
(C) After \citet{VPR+04} (reprinted by permission from Macmillan
  Publishers Ltd.~$\copyright 2004$), graphical representation of the
  macrodomains and their boundaries within the \emph{E~coli}
  genome. Structured macrodomains are indicated as colored arcs, black
  dashed arcs indicate non-structured regions.  (D) From \citet{EMB08}
  (reprinted by permission from John Wiley and Sons
  Ltd.~$\copyright 2008$),
  the genetic insulation of the macrodomains correlates with spatial
  insulation in subcellular territories.  The positions of 10 foci
  were superimposed according to the barycenter of their trajectory
  during 30 intervals of 10 s at the home position. Foci from tags at
  OriC, NSR-3, Right-2 and Ter-6 were plotted.
(E)
Schematic of macrodomains' localization within the cell. Ter is
localized at the periphery of the  nucleoid, towards the cell
membrane~\citet{Meile2011} (reprinted by permission from Biomed
  Central, $\copyright 2011$).
}
  \label{fig:Compartment}
\end{figure}

\begin{figure}[htb]
  \centering
  \includegraphics[width=0.8\textwidth]{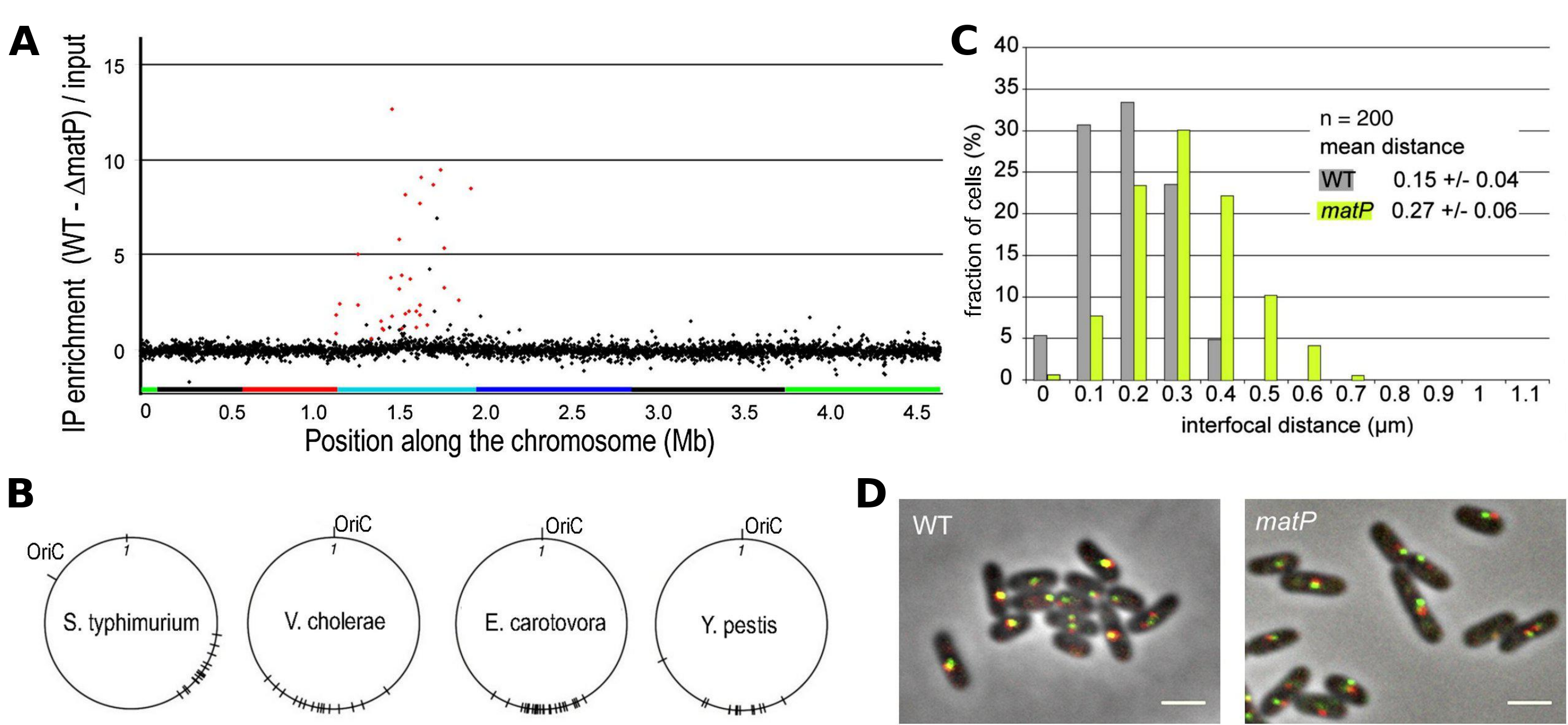}
  \caption{ \emph{The protein MatP organizes the Ter macrodomain by
      specific binding.}  The circular chromosome of \emph{E.~coli}
    divided into 4 macrodomains and 2 non structured zones (see
    figure~\ref{fig:Compartment}). (A) From \cite{Mercier2008},
    ChIP-chip assays for MatP binding show a specific affinity of this
    protein with Ter macrodomains, indicating that the macrodomain is
    condensed by MatP, and suggesting that other macrodomains might be
    condensed by dedicated proteins with a small set of specific
    binding sites.  (B) From \cite{Mercier2008}, the localization bias
    of MatP binding sites is conserved among enterobacteria,
    (\emph{S. typhimurium LT2} with genome size 4683 Kb,
    \emph{E. carotovora SCRI1043} - 5064 Kb), Vibrio
    (\emph{V. cholerae Cl 2} - 961 kbp ), and Pasteurella
    (\emph{Y. pestis KIM 4} - 600 kbp) species.  (C) MatP controls DNA
    compaction in the Ter macrodomain. The histogram represents the
    proportion of cells with given interfocal distances between foci
    of two Ter MD markers in WT and matP deletion mutant cells. (D)
    Fluorescence microscopy images from the two experiments (scale
    bars 2$\mu m$). These results indicate that the effect of MatP on
    foci colocalization is associated with compaction.  (A-D) reprinted
      from \citet{Mercier2008}, $\copyright 2008$, with permission from
      Elsevier.}
  \label{fig:MD}
\end{figure}

Another important recent discovery, consistent with the mentioned
correlation between chromosome arms and cell halfs, is the existence
of ``macrodomains''~\cite{VPR+04,MRB05,EB06} often described as
chromosomal isolated compartments. The first evidence in this
direction~\cite{VPR+04} came from measurements of the recombination
frequency between loci, see figure~\ref{fig:Compartment}A. All else
being equal, this should be proportional to the probability that the
two chromosomal segments come into contact within the cell. For a well
mixed polymer, the recombination frequency should be uniform. However,
experiments show a highly non uniform pattern, compatible with a
compartmentalized structure with clear boundaries.  Four macrodomains
of a few hundred Kb in size have been identified, corresponding to
regions surrounding the replication origin and terminus, and to two
symmetric regions at the edges of the Ter macrodomain, see box and
figure~\ref{fig:Compartment}C,D,E. The remaining ``non-structured regions''
appear to have different physical properties.  Subsequent studies have
confirmed the presence of macrodomains and measured their dynamics
using fluorescently labelled loci~\cite{EMB08,EB06,LMB+05}, see
figure~\ref{fig:Compartment}B,D. While the molecular mechanisms
responsible for this level of organization are not yet clear, the same
authors also found that the Ter macrodomain appears to be condensed by
a single DNA-binding protein (MatP, figure~\ref{fig:MD}) with a small
set of specific binding sites~\cite{Mercier2008}.
Other proteins with macrodomain-specific DNA-binding properties have
recently been
identified~\cite{Dame2011,Sanchez-Romero2010,Tonthat2011,Cho2011}, and
appear to be conserved in related bacteria.
In view of the results from this thread of work, macrodomains might be
seen as a process of microphase separation triggered by specific
protein binding.
Interestingly, as mentioned above macrodomain-like regions also emerge
from independent large-scale genomic
data~\cite{Mathelier2010,Berger2010,Scolari2011}.

During replication, the chromosomes segregate following a well-defined
``choreography,'' which has been the subject of multiple
studies~\cite{Berlatzky2008,Toro2010,Jun2010}.
While segregation is not the main focus here, it is useful to discuss
it briefly, as the existing approaches to this problem (both
experimental and theoretical) are intimately linked with chromosome
organization and will be mentioned in the following.
Specifically, spatial reorganization of the segregating chromosome
arms appears to preserve qualitatively the relationship between loci
distance along the chromosome and in the cell. Moreover,
reorganization ensures that the two replication forks remain in
opposite halves of the cell during replication and that the relative
orientation of the two reorganized nucleoids in pre-division cells is
not random.
Quite interestingly, the spatial separation of sister chromosomes is
not a continuous process, but has been observed to proceed through
``snaps''~\cite{Joshi2011,EMB08}, suggesting the existence of
energetic or entropic barriers for separation, possibly overcome by
active processes.

There is debate on what is the main driver for chromosome segregation:
Entropic repulsion forces due to strong confinement into a box of
linear polymers have been proposed to explain this behavior at least
in part~\cite{Jun2010}; in experiments on
replicating~\textit{B.~subtilis} the chromosome compaction and spatial
organization have been hypothesized to result from non-equilibrium
dynamics~\cite{Berlatzky2008}, rather than from an entropic repulsion
process; in \emph{Caulobacter crescentus,} a contribution from
bidirectional extrusion of the newly synthesized DNA from the
transcription complex has been postulated to contribute to chromosome
segregation~\cite{Jensen2001,Toro2010}.
Experiments using inhibition of protein synthesis by chloramphenicol
show that this produces nucleoids with a more rounded shape and
induces fusion of separate nucleoid bodies~\cite{Helvoort1996}. Upon
release from protein synthesis inhibition, the two nucleoids reoccupy
the DNA-free cell independently of cell
elongation~\cite{Helvoort1998}. The control of the segregation
mechanism by protein synthesis processes could be indirect: ongoing
protein synthesis can affect nucleoid compactness and segregation at
multiple levels, including the decrease in the amount of enzymes
modifying the topology of the DNA or carrying out transcription and
DNA replication.
The idea that membrane protein synthesis activates nucleoid
segregation directly has also been proposed~\cite{Norris1995}  (in
bacteria the presence of cotranscriptional translation creates a
direct physical link between the genome and the membrane~\cite{WN06})
Also note that current arguments based on segregation by entropy and
confinement might turn out to be inconsistent with interpreting
macrodomains as the result of a microphase separation, since a variety
of interactions, and specifically protein-DNA binding processes could
play an important role in defining the free energy of the nucleoid.

\paragraph{Supercoil domains and nucleoid associated proteins.}

\begin{figure}[tbh]
  \centering
  \includegraphics[width=0.8\textwidth]{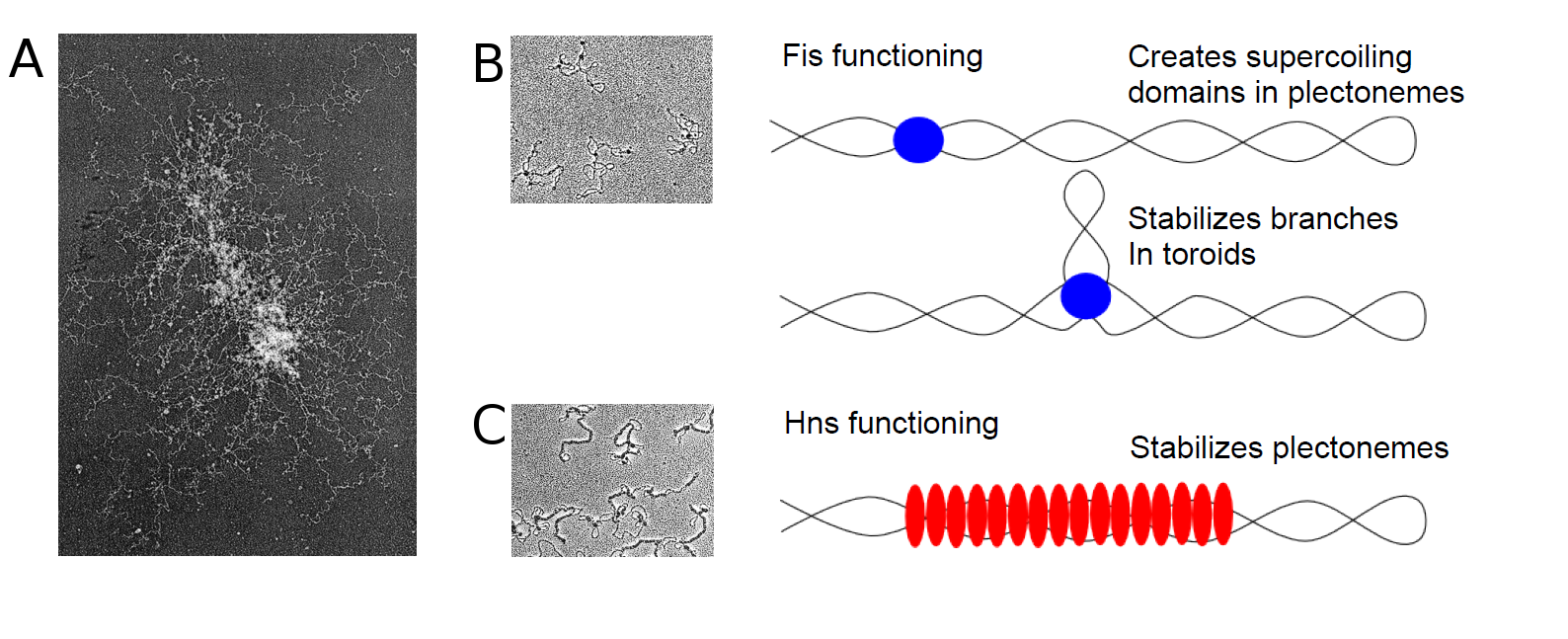}
  \caption{ \emph{Branched plectonemic conformation of the genomic DNA
      molecule in E.~coli and connection with nucleoid associated
      proteins.} (A) From \citet{Postow2004} (Reprinted by
      permission from CSHL Press,~$\copyright 2004$), electron
    micrograph of a purified \emph{E.~coli} chromosome.  The branched
    plectonemic structure is visible. From image analysis, Postow and
    coworkers measured a roughly exponential distribution for the
    length of supercoiling loops, with the average around
    10-15Kb. Note that the extraction and purification process might
    interfere with many of the properties of the conditions in which
    the nucleoid structure is naturally found. (B) Electron micrograph
    of Fis binding to plasmid DNA (from \cite{Schneider2001}).  This
    dimeric NAP stabilizes plectonemic branches of supercoiled DNA
    (schematized in the right panel). In particular, Schneider and
    coworkers measured a Fis to DNA ratio of 1 dimer per 325 bp
    DNA. (C) From \cite{Schneider2001}, electron micrograph of H-NS
    binding to plasmid DNA, elongated complexes (polymers) are formed
    (schematized in the right panel), presumably containing two DNA
    duplexes. Schneider and coworkers measured a ratio of H-NS to DNA
    of 1 molecule per 10 bp. (B-C) reprinted from
      \citet{Schneider2001} by permission of Oxford University Press.
  }
  \label{fig:NAP}
\end{figure}

At smaller scales, the circular chromosome of \emph{E. coli} is
organized in plectonemic loops, or ``supercoil domains.''
Those regions are separated by topological barriers formed by nucleoid
associated proteins (NAPs) such as Fis and H-NS, see
figure~\ref{fig:NAP}. These proteins bridge two strands of DNA by
binding to both, and prevent the propagation of torsional energy. As a
result, they also prevent the spreading of uncontrolled effects on
gene expression in case of accidental DNA breaks or mechanical strain,
caused for example by advancing replication
forks~\cite{Postow2004,Skoko2006,Stavans2006}.

Multiple NAPs have been identified, each with its specific
  binding properties (reviewed in~\cite{LNW+06}). Besides bridging
  double strands, they can change the local shape of DNA inducing
  bends or hinges and form nucleoprotein filaments.
  Additionally, NAPs often have multiple DNA binding modes which might
  be dependent on physiological factors.  A good example of this
  behavior is H-NS. Its binding results in DNA---H-NS---DNA bridges,
  but also forms a rigid nucleoprotein filament which could act as
  zipper \emph{in vitro} \cite{Amit2003,Dame2006,Wiggins2009}.  It
  seems that the nucleoprotein filament formation may be important, as
  it has been found to be a structure shared by other NAPs including
  HU~\cite{vNV+04} and StpA~\cite{Lim2011}.  Biologically, this could
  hypotetically provide a mechanism for environmental sensing by NAPs.
  It is thus possible that our current understanding of NAP binding, being
  based on the limited number of conditions tested, is incomplete and
  generalization of a DNA binding property to other solution
  conditions may be dangerous.

The structure of supercoil domains was studied in \emph{E. coli} by
Postow and coworkers, through analysis of the
supercoiling-sensitive transcription of more than $300$ genes
following relaxation by restriction enzymes \emph{in vivo}, and by
electron microscopy~\cite{Postow2004}.
They concluded that domain barriers may vary dynamically and/or across
a population, but they follow an exponential length distribution. The
average domain size is $\simeq 10-15$ Kb, implying the existence of
about $200-400$ domains~\cite{Postow2004,Stavans2006}. Branches of the
same typical length are visible directly from electron micrographs of
purified nucleoids~\cite{Postow2004,Kavenoff1976}.  Thus, the genome
topology may be visualized as a branched structure with supercoiled domains
that are subject to modulation by nucleoid-associated proteins, and
active processes such as DNA transcription and DNA replication.

The factors responsible for establishing the boundaries of supercoiled
domains and the determinants of domain size and number are still
largely unknown. While H-NS and Fis, along with MukB, the analogue of
the eukaryotic SMC (Structural Maintenance of Chromosomes) protein,
could be involved in stabilizing plectonemic conformations, their
precise roles and importance in this context have not yet been
established in a definitive fashion, either \emph{in vitro} or
\emph{in vivo} \cite{Dillon2010,Skoko2006,Maurer2009,Grainger2006}.

\begin{figure}[htb]
  \centering
  \includegraphics[width=0.95\textwidth]{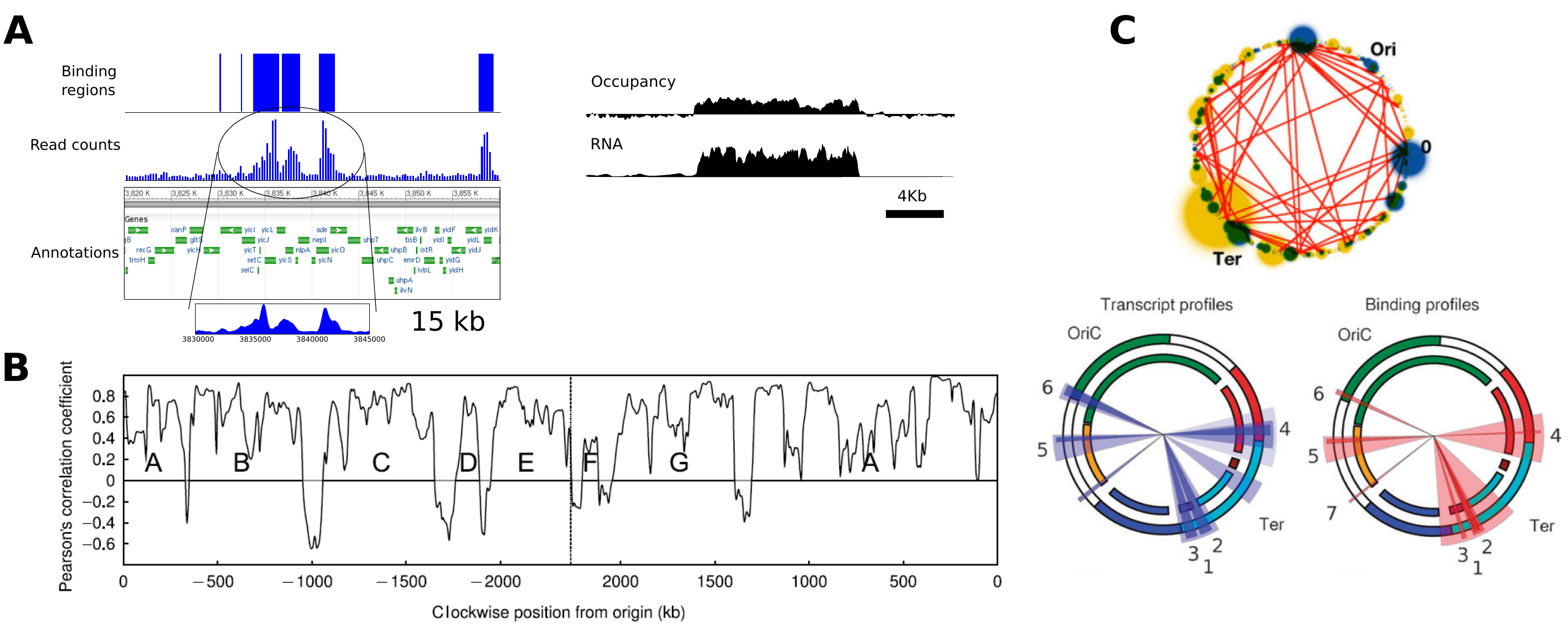}
  \caption{ \emph{Examples of high-throughput data an bioinformatic
      analyses concerning nucleoid organization} (A) Left panel:
    ChIP-seq binding profile of H-NS, plotted using data
      from~\citet{Kahramanoglou2011}; right panel: sketch
      representing the extended protein occupancy domains detected
      by~\citet{vora}. These are polymer-like nucleoprotein
    complexes, presumably related to nucleoid organization. (B)
    Chromosomal sectors emerge from the local correlation (plotted
    here) between expression levels for wild type in log-phase growth
    and a codon bias index, related to translation pressure (from
    \citet{Mathelier2010}, reprinted by permission from Macmillan
      Publishers Ltd: Molecular Systems Biology, $\copyright
      2010$). These sectors correlate well with macrodomain
    organization. (C) Top panel: effective transcriptional regulatory
    network (red links) and areas of influence of Fis and H-NS
    (colored circles) obtained from transcriptomics experiments
    combining NAP mutants and perturbations in supercoiling background
    (from~\citet{MGH+08}, Reprinted by permission from BioMed Central,
    $\copyright 2008$). Bottom panel: clusters of transcriptional
    response to nucleoid perturbations (left, data from~\citet{MGH+08}
    and similar esperiments) and NAP binding (right) correlate with
    the organization of macrodomains (outer arcs) and chromosomal
    segments (inner arcs) of the genome (from~\citet{Scolari2011},
    reproduced by permission of The Royal Society of
      Chemistry). }
  \label{fig:Omics}
\end{figure}

By contrast, a consistent amount of information on the binding of NAPs
in different conditions and its effects on the cell state is available
from high-thoughput experimental techniques (figure~\ref{fig:Omics}.)
Nucleoid associated proteins can modulate the nucleoid conformation
structure in response to changes in environmental
conditions~\cite{LNW+06}. This can result in large-scale changes in
gene expression~\cite{Dillon2010}. The local mechanical action of NAPs
on DNA is often well-characterized by single molecule
experiments~\cite{LNW+06}, which also lead to the observation of
chromatin-like nucleoprotein ``fibers''~\cite{KYH+04}.
Large-scale NAP binding data in specific growth conditions was obtained
from high-throughput experiments involving microarrays (CHiP-chip) or
sequencing (CHiP-
seq)~\cite{Grainger2006,GGL+08,Oshima2006,WSB+07,GAH+07,Kahramanoglou2011}.
Furthermore, transcriptomics studies profiled the changes in gene
expression upon different nucleoid perturbations, such as NAP deletion
and/or altered supercoiling~\cite{blot,MGH+08,bradley,Berger2010}.
Many of these data sets show linear regions of dense binding that
often correspond to macrodomain boundaries, and associate with
global or NAP-dependent transcriptional response and its
correlation with codon bias~\cite{Scolari2011,Mathelier2010}.

The physical origin of this preference in binding and the gene
expression changes at the boundaries of macrodomains is not precisely
clear.  One possibility is that macrodomain boundaries might be
co-localized by NAP structures.
A recent interesting experimental study~\cite{vora} looked at protein
occupancy along the genome regardless of protein identity. This work
uncovered extended polymer-like domains rich in bound proteins
(including NAPs) with an average length of 1.6Kb, associated with
transcriptionally silent or transcriptionally enhanced regions (and
also with high intrinsic DNA curvature.)

When cells enter the stationary phase a radical, global condensation of the
nucleoid occurs. It is believed that this is a mechanism via which the
cell can protect its DNA in harsh conditions~\cite{KYH+04}.
AFM studies have shown that the structure of the DNA differs at the
supercoiling level~\cite{KYH+04} and that action of Dps and CbpA, the
NAPs that replace Fis in this growth phase, is quite different.  The
Dps and CbpA proteins produce compact aggregates (which can protect
DNA from degradation by nucleases) rather than binding to distributed
sites as Fis does~\cite{Cosgriff2010}.  Interestingly, the action of
Fis counters Dps-induced compaction through a transcriptional response
affecting the expression of topoisomerase and gyrase~\cite{OMK+06}.

\paragraph{Compaction by molecular crowding, specific proteins,
  transcription factories, and confinement.}

A different question concerns identifying the main factors
contributing to nucleoid compaction and organization. The main
candidates are macromolecular crowding, electrostatic self-attraction,
supercoiling and nucleoid proteins.

Macromolecular crowding, or the high concentration of macromolecules
present in the cytoplasm, is generally believed to be an important
determinant on the basis of theoretical arguments~\cite{Vries2010,Odi98},
which predict a possible phase separation mechanism between nucleoid
and cytoplasm.  Note that the generic term ``macromolecular crowding''
might include depletion interactions (see below), together with a number of
additional effects of entropic and energetic nature.

DNA condensation by crowders can be observed \emph{in vitro}
  under very controlled and well-understood conditions, for example by
  experiments using dextran or PEG, demonstrating that naked DNA can
  be directly condensed by these
  crowders~\cite{Zhang2009,Estevez-Torres2011,Huang2007,Xu2012}
  Similar (but less controlled) behavior, is shown by purified
  nucleoids~\cite{Zim04}.  At the same time, experimental studies on
isolated nucleoids obtained from mutants lacking various
NAPs~\cite{Zim06b} suggest that the effects of crowding on compaction
are substantial and independent of the NAP composite background.
It has also been suggested that the action of NAPs could be aimed at
antagonizing compaction rather than compacting the
nucleoid~\cite{Zim06b-a}.
However, this must be a complicated, combined effect involving forces
of different nature. In the absence of crowding and confinement, it is
obvious that the radius of gyration of the genome would be smaller if
it were organized, e.g. in a branched structure of plectonemic loops
stabilized by DNA-bridging NAPs~\cite{Postow2004,Trun1998}.


As a particular case of crowding effect, it has been proposed that the
(attractive) depletion interactions, well known in colloid science,
might play an important role in chromosome
organization~\cite{Marenduzzo2006}
This force is due to reduction in total solvent excluded volume upon
formation of a molecular complex.
Depletion interactions are consequential when large molecular
assemblies are formed in presence of smaller particles. In fast-growth
conditions, genome-bound RNA polymerase is localized into a few
transcriptionally active foci or ``transcription
factories,''~\cite{JC06,Cabrera2009}, analogous to the eukaryotic
case~\cite{Marenduzzo2006,Marenduzzo2006c}.
Depletion interactions were suggested to explain the formation of
these macromolecular assemblies~\cite{Marenduzzo2006c}. Interestingly,
the formation of these foci has been associated with the presence of
the NAP protein HU~\cite{Berger2010}.
We add that a similar argument might hold for the local compaction of
the surroundings of OriC, rich in ribosomal RNA producing regions, in
a macrodomain-microphase.
In other words, ribosome-rich ribosomal RNA transcripts, attached to
the genome through RNA polymerase, could help compact the Ori region.
If this should be the case, the compaction properties of this region
would change with the number of ribosomes being synthesized, i.e. with
growth rate and translation efficiency~\cite{Scott2011} (One can also
speculate that, for the same reason, newly-replicated DNA could be
sequentially aggregated during replication.)
On one hand, experiments show that transcription of ribosomal RNA
operons (which are generally located in the Ori macrodomain) is
related to the compaction of nucleoids observed upon inhibition of
translation~\cite{Cabrera2009}.
On the other hand, these processes must be intersected with the (non
entropic) binding of NAPs, given the evidence connecting specific NAPs
to the compaction of the Ter macrodomain (MatP) and indirectly to the
effect of the HU protein on organization of the Ori macrodomain.

\begin{figure}
  \centering
  \includegraphics[width=0.8\textwidth]{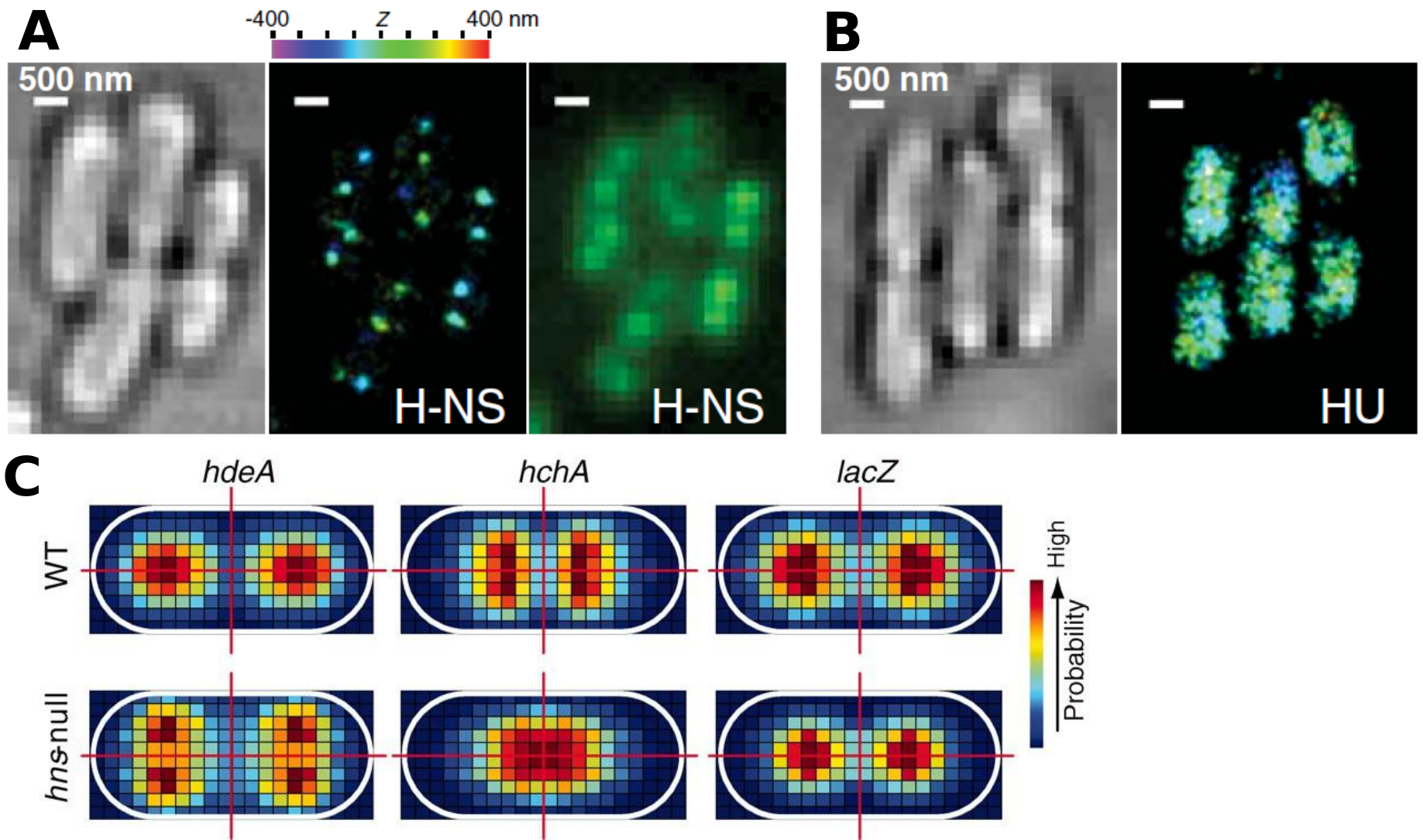}
  \caption{\emph{Effect of H-NS on chromosome compaction.}
    From~\citet{Wang2011a}, Reprinted with permission from
      AAAS. H-NS was shown by Wang and coworkers to form a small set
    of foci in the cell, which bring together distant H-NS
    targets. Together with MatP (figure~\ref{fig:MD}), this evidence
    proves a role of drives of energetic origin in chromosome
    compaction. (A) Super-resolution fluorescence imaging of
    fluorescent protein fusion of H-NS showing compact H-NS clusters
    in the nucleoid. The \emph{E. coli} cells are shown in the
    bright-field image (left). The z coordinate of each localization
    is color-coded (top bar). In comparison, a conventional
    fluorescence image of the same cells is shown (right).  (B)
    Scattered distribution of HU in the nucleoid. (Left) Bright-field
    image; (right) super-resolution image. Similar diffuse
    distributions were observed for Fis, IHF, and StpA.  (C) Effect of
    deleting H-NS on the subcellular distribution of H-NS target genes
    (hdeA, hchA) versus non-targets (lacZ). The 2D histograms of the
    relative hdeA, hchA, and lacZ locus positions normalized to the
    cell dimensions are shown, the genes were fluorescently labelled.
    The first quartile of the cell plots the (color-coded)
    probabilities of finding the fluorescent locus in the particular
    position, the rest of the cell is filled with mirror images to aid
    the eye (grid size 100 to 200 nm). In each case, 2000 to 5000 gene
    locus positions were analyzed. H-NS deletion has little effect on
    lacZ distribution, despite this locus is only a few Kb apart from
    the closest H-NS binding site.}
  \label{fig:HNS}
\end{figure}

 A very recent
study~\cite{Wang2011a} systematically addressed the chromosomal
localization and role in spatial organization of the nucleoid of five
major NAPs (HU, Fis, IHF, Stpa, H-NS) using fluorescent protein
fusions and super-resolution fluorescence microscopy. In the growth
conditions tested
all the proteins showed scattered distributions in the nucleoid,
except for H-NS, which seemed to form (on average) two distinct foci
per chromosome copy, bringing together different (even distant) H-NS
targets, see figure~\ref{fig:HNS}. Thus, H-NS should be added to the
list of NAPs with a compacting action on the nucleoid associated with
the formation of specific foci. The long-range interactions between
H-NS binding targets were validated by 3C, and show no apparent
coherence with the macrodomain structure; loci pairs that are near to
H-NS targets but are not targets show no 3C signal and wider
distributions of subcellular distances (evaluated with microscopy).
In general, DNA associated with many NAPs has a much larger
  surface, which should enhance the depletion interactions. For
  instance, DNA coated with an H-NS nucleoprotein filament will have a
  diameter of about 20 nm instead of the 2 nm of naked DNA.  H-NS
  nucleoprotein filament formation could strongly enhance the
  depletion attractions with respect to other NAPs not forming
  filaments, and filaments formed at remote locations on the contour of
  the chromosomal DNA could find each other in a crowded
  environment. This may explain the results from super-resolution
  imaging mentioned above~\cite{Wang2011a}.
  We note again that this is one of the counter-intuitive aspects of
  molecular crowding, since one might think that it would be easier to
  encounter another filament if there were no ``obstacles''. The
  confusion can come from thinking about diffusion, which is hindered
  in a crowded environment due to the higher effective viscosity,
  versus thermodynamics, where the depletion interactions increase the
  probability of observing two nucleofilaments in contact at
  equilibrium. Adopting a simple Kramers-type model for the kinetics,
  it is not obvious whether the rate of association would go down due
  to the prefactor or go up due to the lower free energy minimum. Implicit
  in the entire discussion implicating crowding in aggregation in this
  cellular context is the idea that the distribution of nucleoprotein
  filaments is at near-equilibrium even though the cell is dividing
  rapidly.

Finally, depending on the degree of autonomous compaction, the
confinement exerted by the cell wall might play a relevant role in
nucleoid organization and segregation~\cite{Jun2010}.
This is expected to be particularly significant in fast-growth
conditions, where the genome needs to be highly accessible for
transcription (and thus will not be condensed) and the amount of
genome per cell is higher due to overlapping replication
rounds~\cite{NYH+07}.

To summarize, the degree of condensation and the geometry of the
nucleoid are strongly dependent on the growth phase and growth rate of
the bacteria. Such changes may be modeled by means of equilibrium
states, slowly evolving in accordance to ``external'' control
parameters (e.g. the concentration of the various NAPs).
However, it seems likely that nonequilibrium processes are also important,
and relevant aspects of the problem might be lost in attempting to
describe the nucleoid condensation process purely in terms of an
approach to thermodynamic equilibrium.
%
%
For example, on the other side of the spectrum in terms of biological
complexity, the recently found scale-invariant structure of the human
genome~\cite{Mirny2011,Lieberman-Aiden2009} was suggested not to be
the result of an equilibrium state, but similar in nature to the
so-called ``crumpled globule,'' or ``fractal
globule,''~\cite{Grosberg1993} since loci that are near along the
genome arclength coordinate are also physically proximal in
three-dimensional space.
This proximity contrasts with what happens in an (equilibrium)
collapsed polymer, where the linear structure is completely mixed in
the globule.

Let us quote a few experimental findings, some of which are very
recent, supporting the role of nonequilibrium processes, specifically
for bacteria.
The initial conditions, i.e. the choreography of DNA replication, appear
to play a central role in defining the final structure of the
nucleoid~\cite{Daube2010}.
In artificial \emph{E.~coli} strains with two distant origins instead
of just one~\cite{Wang2011}, the two origins initiate replication
synchronously at the expected separate positions of the genetic loci
associated with them. Replication forks move independently, indicating
that replication does not occur in a single replication factory and
that the replication machinery is recruited to origins rather than
 vice versa. Most importantly, in these experiments
progression of replication plays a major role in determining the
space-time pattern of locus segregation.
The large scale structure of the \emph{B.~subtilis}
nucleoid~\cite{Berlatzky2008} has been observed at various stages of
the replication process. The newborn portions of the chain are
compacted and sequentially conveyed towards the poles, resulting in an
ordered, spiraling structure. A strong correlation between space
coordinate and genomic coordinate is preserved, similar to the linear
behavior observed in \emph{E.~coli}~\cite{Wiggins2010}.  A
choreography of this sort is found also in
\emph{Caulobacter}~\cite{Jensen2001}. Finally, the large scale spiral
structure of the nucleoid of \emph{Bdellovibrio
  bacteriovorous}~\cite{Butan2011} also suggests a metastable steady
state, sustained by cooperative motion and/or energy exchanges. It
seems difficult to disregard the deterministic replication-segregation
dynamics in describing such phenomenology~\cite{Breier2004,Toro2010}.

\begin{figure}[t!bh]
  \centering
  \includegraphics[width=0.8\textwidth]{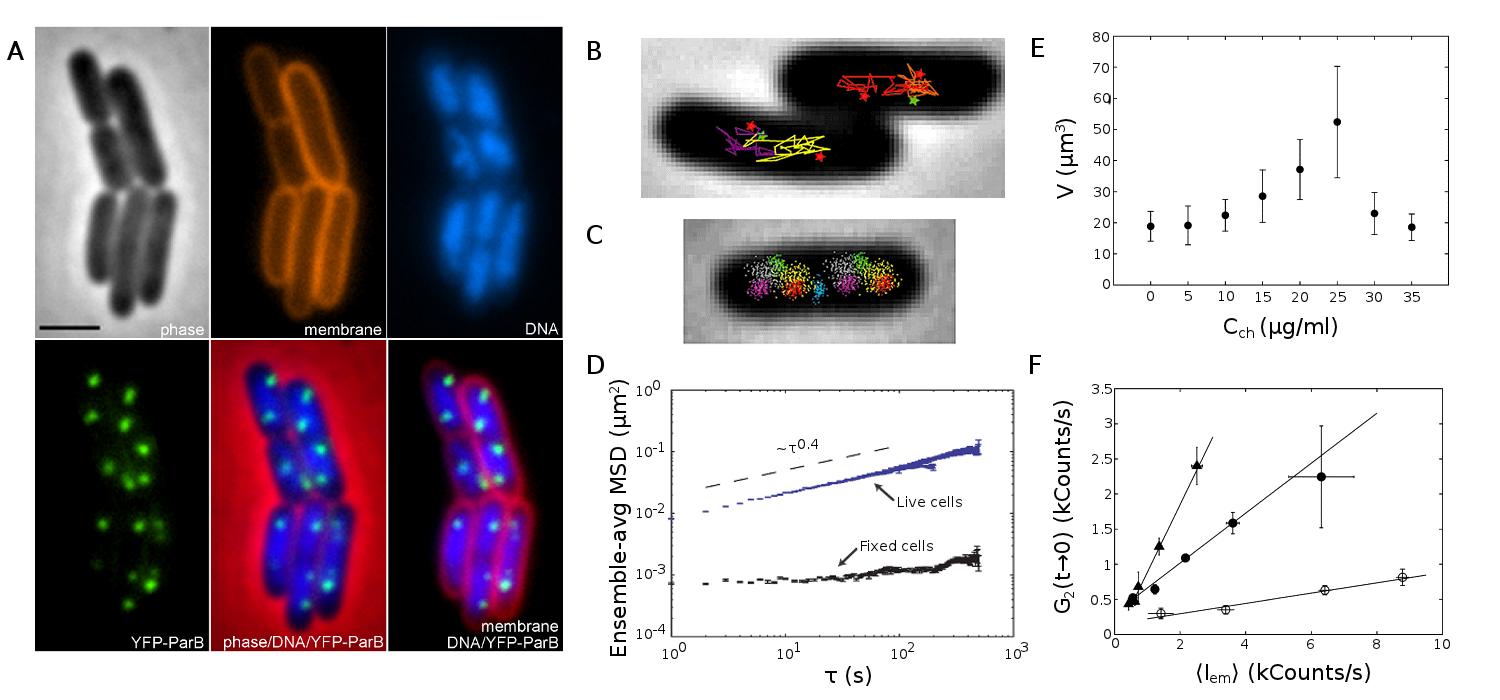}
  \caption{ \emph{Optical techniques in nucleoid dynamic
      visualization.} (A) From \citet{Meile2011} (reprinted by
      permission from Biomed Central, $\copyright 2011$), using phase
    contrast for cell visualization, membrane staining (FM 4-64), DNA
    staining (DAPI), and YFP-ParB for foci dynamic measurement, it is
    possible to produce ``informed maps'' of the bacteria by
    overlaying different micrographs.  (B) Movies can be used to to
    follow foci dynamics.  The plot, from \citet{EMB08} (
      reprinted by permission from John Wiley and Sons
      Ltd.~$\copyright 2008$), shows a typical trajectory of a
    fluorescent focus, from which dynamical properties of the nucleoid
    can be inferred. (C) From \citet{EMB08},  reprinted by
      permission from John Wiley and Sons Ltd.~$\copyright 2008$. The
    trajectories of fluorescent dots define territories in which
    genetic loci are placed within the cell.  (D) From
      \citet{Weber2010}
      (http://prl.aps.org/abstract/PRL/v104/i23/e238102),
      reprinted, $\copyright 2010$ by the American Physical Society.
    A measurement of the ensemble-averaged MSD for live and fixed
    cells using fluorescent loci.  The authors showed that the
    dynamical exponent of the foci is universal (and close to
    $0.4$). The observed anomalous diffusion has been explained
    phenomenologically as a consequence of the viscoelastic behaviour
    of the nucleoid. We hypothesize that this could be a consequence
    of a fractal organization of the nucleoid similar to a ``fractal
    globule'', \emph{see Discussion}. (E, F) Measurements on purified
    nucleoids, from \citet{RFK07}, reprinted with permission from
      Elsevier, $\copyright 2007$. Using Fluorescence Correlation
    Spectroscopy, nucleoid volume and amplitude of FCS correlation
    functions have been measured at varying supercoiling (induced by
    different concentrations of chloroquine drug). The plots show that
    the nucleoid reach a maximum volume and minimum FCS correlation at
    full supercoiling relaxation. The FCS correlation signal can be
    used to deduce the size of structural units of the polymer.}
  \label{fig:DIN}
\end{figure}

\paragraph{Viscoelasticity and structural units. }
Tracking studies of fluorescently labeled chromosomal loci have
evaluated \emph{in vivo} dynamic properties of the nucleoid with
fairly high time resolution, measuring for example the mean-square
displacement (MSD) of the loci or the time autocorrelation
function~\cite{EMB08,Weber2010,Meile2011}. In general, these
measurements give information on the local relaxation time scales of
the nucleoid and its viscoelastic behavior, see figure~\ref{fig:DIN}.
On one hand, for large time-scales~\cite{EMB08}, loci mobility
correlates well with macrodomain structure. In particular, the MSD
saturates at the spatial scale of the macrodomain size. On the other
hand, especially for smaller time scales, the mean square displacement
of a locus is seen to follow a power law: $MSD \equiv <(\vec{x}(s,t+
\Delta t)-\vec{x}(s,t))^{2}> \approx c \cdot (\Delta t)^{\alpha}$
(where $s$ is its  arclength genomic coordinate) and the
exponent $\alpha$ seems to be universally close to $0.4$,
  independent of $s$~\cite{Weber2010}. Perhaps surprisingly, an
extra-chromosomal RK-2 plasmid showed the same behavior, while smaller
RNA particles had a higher subdiffusive exponent. It must be mentioned
that the localization of RK-2 plasmids appears to be highly
regulated~\cite{Kolatka2008,Derman2008}.
The underlying viscoelasticity of the bacterial cytoplasm surrounding
the nucleoid is still poorly understood. Some characterisation has
been approached \emph{in vivo} via FRAP measurements of diffusing
GFP~\cite{Konopka2006}.


The observed anomalous diffusion has been modeled using
phenomenological approaches. There are a variety of dynamical models
exhibiting anomalous diffusion, such as Langevin equations with time
dependent viscosity (also equivalent to fractional Langevin
equations), continuous time random walks, and  random walks over a
fractal object~\cite{Burov2008,He2008,Condamin2008}.  Weber et
al.~\cite{Weber2010,Weber2010b} compared their data with the results
obtained on the basis of the first two approaches, and the fractional
Langevin equation gave a more satisfactory agreement.

The anomalous diffusion exponent \emph{per se} does not give
quantitative information on the geometry of the nucleoid.  Physical
measurements, possible on purified
nucleoids~\cite{COS+01,Cunha2005,RFK07}, can enrich the scenario. In
particular, Romantsov and coworkers~\cite{RFK07} have obtained, using
fluorescence correlation spectroscopy, the time-dependent
coarse-grained density distribution of purified nucleoids. They
conclude that the polymer appears to be composed of a set of
``structural units'' defined by a measurable correlation length.  The
measurements were performed at varying  degrees of supercoiling, induced by
different concentrations of the chloroquine drug. The size of the
structural units was found to vary from 50-100Kb in high (positive or
negative) supercoiling to 3Kb at zero supercoiling. The diameter of
the purified nucleoid varied from 2.5$\mu$m in high supercoiling to
3.5$\mu$m in low supercoiling.
The authors also estimated the typical diameter of the structural
units from the diffusion constant (obtained from the decay of the
fluorescence autocorrelation function) and Stokes-Einstein's
relation. Perhaps surprisingly, the resulting size of structural units
was near 70-80 nm regardless of supercoiling.
Thus the emerging picture for the unconfined genome is that of a
string of $\sim 100$ highly dense ``beads'', each containing $\sim
100$ Kuhn lengths (effective independently jointed elementary polymer
segments) of DNA each. These values apply in the presence of
supercoiling but in the absence of crowding and confinement effects,
as for purified nucleoids most of the cytoplasmic (and probably a
considerable part of the DNA-binding) proteins are probably diluted
away.
%

\section{Models}

We will now review a few modeling approaches put forward in  recent
years, and point to some more classic work in polymer physics that we
believe could be relevant in this context.
As the reader will have observed, the wealth of existing experimental
results is appealing on one hand, but, on the other hand, does not
offer any clear grasp on a small set of relevant ingredients necessary
for building coherent physical descriptions of the
nucleoid. Consequently, the approaches adopted in the literature are
highly diverse and heterogenous in terms of premises, methods,
ingredients, and points of view.

The notable efforts to understand entropic aspects of chromosome
segregation (reviewed in \cite{Jun2010}) have revived the studies on
entropic forces of single and multiple confined polymers dating back
to Edwards and DeGennes~\cite{JM06,DeGennes1979,Edwards1969,Jung2010}.
As we have pointed out above, the role of bound proteins on nucleoid
entropy is typically disregarded in these arguments.

A comprehensive review of the literature on confined DNA in different
contexts is provided in~\citet{Marenduzzo2010}.
Considering this approach, there remains the open issue that the
observed segregation and compaction times might be too short to be
compatible with an entropic process. For linear polymers this time has
been evaluated by Molecular Dynamics simulations to scale like the
square of the number of monomers, $N^2$, which, for sufficiently large
$N$, will be smaller than the $O(N^3)$ chain diffusion
time~\cite{Arnold2007a}.
%
However, it is not straightforward to use these scaling relations for
empirically relevant estimates.  Moreover, these estimates will depend
on the model used to represent the structure of the nucleoid and its
correlation with the replication process, linking this problem to
other unanswered questions.
Overall, it seems likely that entropy is only part of the
story, and active / nonequilibrium processes of different kinds might
play a role in chromosome segregation.

We will now turn our attention to work concerning
nucleoid organization and cellular arrangement.
Buenemann and Lenz have attempted to understand the linear arrangement
of chromosomal loci in terms of a purely geometrical
model~\cite{Buenemann2010}, where a linear polymer in the form of a
string of blobs, is confined within a cylinder and locked at one or at
a few loci.
This constrained geometry obviously provides an ordering mechanism, as
long as the blobs are large enough with respect to the cylinder's
diameter. This model makes predictions on the spatial arrangement of
the chromosome in mutants of \emph{C. crescentus}~\cite{Viollier2004a}
and on the cell-cycle dependent ordering in \emph{E.~coli}.
A very recent simulation study~\cite{Fritsche2011} explains the linear
ordering observed in \emph{E.~coli} as a product of confinement and
entropic repulsion of a string of linearly arranged chromosomal
loops. In order to show this, they represent the chromosome as a
confined circular self-avoiding chain under the constraint that
consecutive loops, identical in size, are distributed along the
arclength coordinate, while the Ter region does not contain such
loops. Their simulations show both linear ordering along the cell axis
and Ter region occupying the outer periphery of the
nucleoid~\cite{Meile2011}, as properties of the equilibrium states.
Intriguingly, the same study suggests that this linear-loop ordering
could be a consequence of the transcription network organization. To
support this point, they simulate a polymer where transcription
factor-target pairs are coupled by attractive harmonic interactions,
and show that the linear ordering is recovered. It is well-known that
 transcription factor-target distances have a statistical tendency to be
short along the chromosome~\cite{Warren2004}.

\begin{figure}[tbh]
  \centering
  \includegraphics[width=0.8\textwidth]{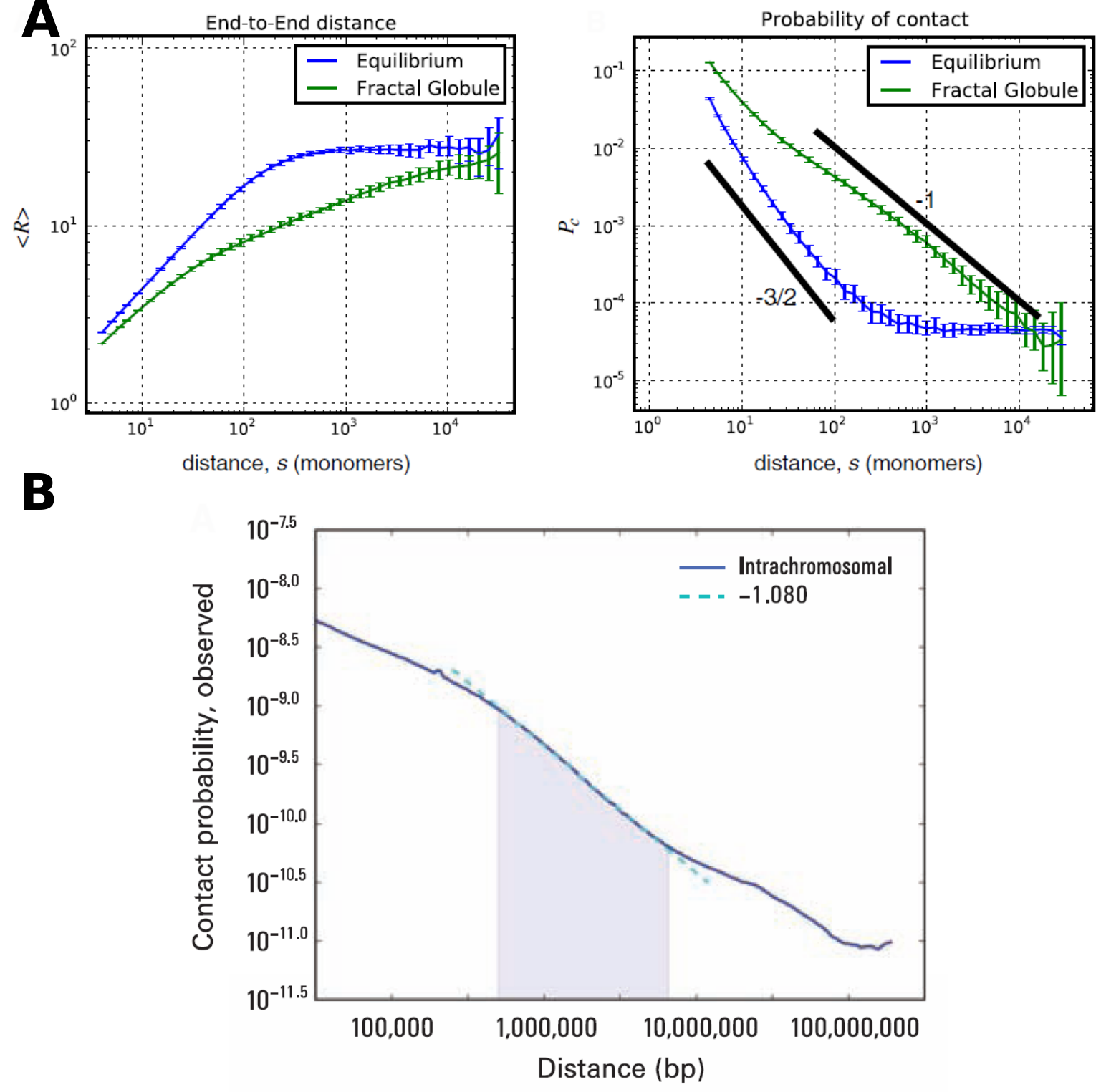}
  \caption{ \emph{Main universal features of the ``fractal globule''
      model}.  (A) From \citet{Mirny2011} (with kind permission
      from Springer Science$+$Business Media), root-mean squared
    end-to-end distance $R(s)$ and probability of a contact $P(s) $ as
    a function of the genomic distance $s$ between the ends of a
    subchain of $N=32000$ monomers, from simulations of a fractal
    globule and a collapsed (equilibrium) globule.  Notice the
    different scalings in the two situations.  All random collapsed
    configurations of a polymer in two dimensions behave as a fractal
    globule, as topological entanglement is forbidden. This does not
    hold in three dimensions, as a large number of knots can be
    generated. In a kinetically constrained situation where knots are
    not present, this property is restored, and the resulting
    metastable configuration has the property that segments that are
    close in arclength distance, are also close in space.  It has been
    argued that this configuration might be relevant for eukaryotic
    chromatin within a range of length and time scales.  (B) From
    \citet{Lieberman-Aiden2009} (reprinted with permission from
      AAAS), contact probability as a function of genomic distance
    averaged across the genome (blue) of an eukaryotic cell obtained
    by the Hi-C technique. It shows a power law scaling between 500 kb
    and 7 Mb (shaded region) with a slope of –1.08 (fit shown in
    cyan). This analysis indicates that the eukaryotic DNA might be
    organized similarly to a fractal globule.  }
  \label{fig:FGL}
\end{figure}

Vettorel and coworkers~\cite{Vettorel2009} performed an abstract study
motivated by the possible nature of the compartmentalization and the
structural units of a generic (eukaryotic or prokaryotic) chromosome
forming a crumpled or fractal globule mentioned in section~\ref{sec:meas}
(figure~\ref{fig:FGL}). Specifically, they explore metastable
collapsed states of polymers, where the total size scales as $N^{1/3}$
as in an equilibrium compacted globule, but a much higher degree of
compartmentalization is present.
At odds with the intrinsic disorder of the equilibrium globule, in a
fractal globule, for any pair of loci separated by a chain length $s$,
the distance $R(s)$ has the scaling behavior $R(s)\approx
s^{1/3}$~\cite{Mirny2011}.
In other words, both the equilibrium globule (i.e. the equilibrium
collapsed structure emerging from polymer self-attraction or
unfavorable entropy of mixing) and the fractal globule have mass
fractal dimension $D_{f}=3$, but a generic volume (and in particular a
blob in the DeGennes sense) inside the equilibrium globule can include
non-sequential segments, while in the fractal globule it contains a
single sequential segment~\cite{Grosberg1993}.
%
This structure can be understood as resulting from a process where
condensation sequentially involves larger and larger scales in $s$, so
that the genomic proximity is preserved in a scale invariant fashion.
As a consequence, distant portions of the chromosome will occupy
different compartments within the globule.

In order to obtain a fractal globule in a simulation, a constraint
preventing entanglement must exist~\cite{Vettorel2009}. To obtain this
condition, Vettorel and coworkers considered a semi-dilute or
concentrated solution of mutually disentangled (unconcatenated) rings.
Also note that because of the constraint preventing entanglement
during collapse, a fractal globule is generally larger in size than an
equilibrium globule of equal chain length (while for both the total
size scales as $R \approx N^{1/3}$).
The fractal globule configuration has proven to be relevant for the
description of genomic DNA organization in non-dividing eukaryotic
(human) cells~\cite{Lieberman-Aiden2009,Mirny2011}.
In particular, it has been found that the human chromosomes display
this structure from $\Delta s \approx 500$Kb to $\Delta s \approx
7$Mb. This range of sizes might be relevant to bacteria that have
genomes spanning a few Mb (also because the lower cutoff might be
related to the fiber organization of chromatin, which in bacteria is
different).  Thus, as the authors speculate, the fractal globule
description might be useful for the nucleoid as well. We will further
discuss this point in the light of the available data in
section~\ref{sec:Disc}.

We have already mentioned above the theoretical work on the role of
macromolecular crowding on compaction~\cite{Vries2010,Odi98}, and of
depletion interactions on loop formation (and possibly on macrodomain
organization~\cite{Marenduzzo2006c}).
A recent simulation study~\cite{Junier2010} has concentrated on the
chromosome-shaping role of transcription factories. Considering a
self-avoiding worm-like chain with a fixed hard-core repulsion radius,
and short-ranged bridging protein complexes, they show that this
system can take ``micro-structured'' collapsed globule configurations,
where bridging complexes cluster and regions of high and low densities
of interacting sites coexist in a microphase separated thermodynamic
state.
More abstract analytical studies could provide a useful context for
understanding the role of NAP binding~\cite{DA00,Kantor1996a}.

Finally, some attention has been devoted recently to the role played
by the branched plectonemic structure of the
nucleoid~\cite{Odijk1998,Ubbink1999}, which is believed to have
relevant implications for transcription~\cite{Dillon2010,Postow2004}.
Provided the correct questions are formulated, this topic has
the advantage of being placed in a strong framework developed in the
past 30 years, building on the classical calculations of Zimm and
Stockmayer~\cite{Zimm1949} for considering arbitrarily ramified ghost
chains as gaussian networks, and obtaining their equilibrium
properties~\cite{Sommer1995,Farago2000a,Graessley1980}. For example,
some recent work has focused on the induction of loops involving
multiple polymer segments~\cite{Sumedha2008}.
In this framework, reliable estimates for the self-avoiding case can
be obtained with Flory-like arguments, taking into account the fact
that branched polymers have higher internal topological complexity,
which makes the repulsive interactions stronger than between linear or
circular chains.  The same approach is also possibly relevant for
studying dynamic aspects of loci mobility
~\cite{Jasch2003,Dolgushev2009}, which can be compared with
simulations and more phenomenological models~\cite{Weber2010b}.
Finally, a number of very refined results based on renormalization
group / field theoretical methods have been obtained in a more
abstract context using the ``randomly branched polymers'' model (or
``lattice animals''), typically defined as the ensemble of all the
clusters of connected sites (monomers) on a regular lattice.
While this ensemble is probably too general, it is possible that these
results have implications for questions related to the structure of
the nucleoid.
For example, recent work based on Langevin dynamics~\cite{Janssen2011}
analyzes their collapsed regime, obtaining a fractal dimension
$D_{f}=2.52$, intermediate between the swollen chain $(D_{f}=2)$ and
the fully compacted globular state $(D_{f} =3)$.

%


\section{Discussion. Hypotheses and paradoxes concerning nucleoid
  geometry and dynamics. }

\label{sec:Disc}

We would like to discuss here some speculations on the possible links
between the experimental and theoretical results discussed above.

Let us start with an analogy to the geometrical organization of
eukaryotic chromatin, where different geometrical features are
observed at different scales, ranging from the known fiber
organization up to the arrangement in preferred ``chromosome
territories'' within the nucleus.
In this case, experimental techniques such as Chromosome Conformation
Capture (3C) and its high-throughput variants
\cite{Lieberman-Aiden2009}, or FISH (fluorescence in situ
hybridisation, where fluorescent tags are attached to pairs of DNA
loci) allow, for instance, to measure the mean absolute distance, $R$,
and the contact probability, $P_c$, of two genomic loci at arclength
distance $s$. For bacteria, these techniques entail a number of
specific difficulties, but the data are starting to be available, as
already mentioned~\cite{Umbarger2011}. However, the study by Umbarger
and coworkers focuses on the large-scale nucleoid 3D architecture,
rather than on more detailed properties of the interaction map.
Apart from saturation at large genomic
distances, $R(s)$ is typically found to increase with
$s$~\cite{Mateos-Langerak2009} with an approximate power-law behavior,
$R(s) \sim s^{\nu}$.
Here, $\nu$ is a scaling exponent, interpretable as $1/D_f$, which
empirically varies with scale of observation and cell type in
eukaryotes.
If a stretch of length $s$ of a polymer spans a region of size $R$, in
$D$ dimensions, it can occupy a volume $V \sim R^{D} \sim s^{\nu D}$.
Thus, heuristically, one expects that the probability of one end of
the stretch meeting the other scales as $P_c \sim 1/V \sim s^{- D
  \nu}$.
%
Experimentally, for loci on the same chromosome, $P_c(s)$ decreases as
a power-law $P_c(s) \sim s^{-1}$, for a set of length scales in the
approximate interval 0.5-7Mb for $s$~\cite{Lieberman-Aiden2009}, which
provides evidence for a fractal globule-like organization, $\nu =
1/3$. This is confirmed by FISH data for $R(s)$ on a smaller range of
genomic lengths~\cite{Mateos-Langerak2009}.

However, experimental data on $R(s)$ for chromatin are complex, as the
$\nu$ exponent is cell-type specific and varies with genomic length,
reflecting different degrees and modes of chromatin compaction.
At short genomic distances $\nu$ is found in the range $0.2-0.6$ at
short distances, and $R(s)$ reaches a plateau (i.e., $\nu\sim 0$) at
order 10 Mb genomic distances, because of chromosome
territories~\cite{Shopland2006,Mateos-Langerak2009,Nicodemi2011}.
For the \emph{E.~coli} nucleoid, the only available quantitative
data~\cite{Wiggins2010} indicate that $R(s)$ might scale like $s$ for
non-replicating chromosomes at scales of 0.3-2 Mb.

In principle, a link between nucleoid geometry and the measured
nucleoid local dynamics is expected.
The anomalous diffusion of chromosomal loci in \emph{E~.coli} has been
modeled in terms of fractional Langevin equations~\cite{Weber2010b};
such an approach correctly reproduces the temporal behavior of the
loci, but disregards the geometry of the structure containing them.
Very likely, for any polymer, this structure has a fractal character,
and one would like to understand how this influences the motion of the
loci.  A minimal model would consider a relaxation equation over
the fractal.  For a purely self-avoiding polymer one has $D_{f}
\approx 1.7$. In the case of a self-attracting polymer where
attraction is screened by self-avoidance, i.e. at the $\theta$
point~\cite{DeGennes1979,Grosberg1994}, it is reasonable to assume a
mass fractal dimension $D_{f}=2$, as for a ghost chain.  When the
tendency towards compaction increases, one expects that $D_{f}$ will
increase accordingly.

In order to illustrate this point with an example, we can consider
protein structures.  Data from the Protein Data
Bank~\cite{Berman2000} for 200 proteins with a number of amino acids
ranging from $N \approx 100$ to  $N \approx 10000$ give values of
$D_{f}$ from 2.3 up to 2.6 ~\cite{Enright2005}.  On a larger scale,
high resolution X-ray spectroscopy has resolved the ribosome
structure at the atomic level. Such data indicate~\cite{Lee2006} that
the heavier  50S unit is fully compacted, with $D_{f}= 3$, while the
lighter 30S unit has $D_{f}= 2.8$; it has been argued that the sparser
structure of 30S is compatible with a dynamic geometry, as required in
the translation process.
Folded proteins are generally described as an harmonic network, by the
so-called Gaussian Network Model~\cite{Reuveni2010}, and we can try to
apply a similar reasoning to the nucleoid.
Close to the fully collapsed regime of a very long polymer such as a
bacterial chromosome, where one expects $D_{f} \approx 3$, the
harmonic approximation strictly does not apply, because hard-core
  repulsion acts against chain compression.
The total energy can then be written as an harmonic ``Rouse'' term,
describing waves propagating along the chain, plus a
  self-attraction term and a self-avoidance term.
The relaxation dynamics in such conditions (neglecting hydrodynamic
interactions) has been studied within a continuum model approach by
Pitard and Orland~\cite{Pitard1998}. They find that the relaxation
time $\tau$ of the globule scales with the polymer length $N$ as $
\tau \approx N^{5/3}$.  Taking into account that the globule size $R$
scales as $R \approx N^{1/3}$ (i.e. it fills space),
this implies that $R^{2} \approx (\tau^{3/5})^{2/3} \approx
(\tau)^{2/5}$ (since $N \approx \tau^{3/5}$).  This scaling appears to
coincide with the experimental value for the anomalous diffusion
exponent $\alpha = 0.4$ measured by Weber and
coworkers~\cite{Weber2010}, previously quoted in the text.
In other words, the Rouse subdiffusive dynamics of a collapsing (and
thus off-equilibrium) globule follows the same scaling law as the observed
local dynamics of nucleoid loci within a range of time scales.
While this might be simply a coincidence, it leads us to speculate
that the measured dynamic exponent for the mean-square displacement
might be the consequence of a fractal-globule-like nature of the
nucleoid, at least within a range of length scales.

This argument can be recast in more generic terms.  Rouse polymer
relaxation dynamics was originally explored by
De~Gennes~\cite{DeGennes1976}, who obtained the scaling relation $z= 2
+ D_{f}$, where $z \equiv 2/\alpha$ is the so-called dynamical
exponent.  At the $\theta$ point~\cite{DeGennes1979,Grosberg1994}
where one has $D_{f}=2$, the DeGennes's relation gives $\alpha=1/2$,
the Rouse result for non-interacting chains.  As the polymer dimension
$D_{f}$ increases, a smaller value of $\alpha$ is to be expected; in
the compacted configuration, where $D_{f}=3$, the relation gives
$z=5$, which is the result reported above.
DeGennes's work is based on scaling arguments, but is confirmed by
field-theoretical methods~\cite{Wiese1998}
for two-body interactions.

To our knowledge, in the collapsed regime the relation has been proved
only at the level of mean field~\cite{Pitard1998} by modeling, in the
spirit of a virial expansion, the effective interaction with an attractive two-body term and a repulsive
three-body term.


In conclusion, if the nucleoid behaves like a fractal globule (i.e. an
off-equilibrium polymer collapsing because of self attraction, but
where entanglement is prevented by topological constraints), or more
in general if it has fractal dimension $D_{f}=3$, from mean field
theory one expects the subdiffusion exponent $\alpha=0.4$.
Conversely, if the DeGennes relation is valid, the experimental result
$\alpha \approx 0.4$, observed in \emph{E.~coli} and in large
plasmids~\cite{Weber2010}, implies that the nucleoid fractal dimension
could be $D_{f} \approx 3$.

To our knowledge the available experimental result that comes closer
to a direct measurement of $D_f$ deals with the mean square
displacement of fluorescently tagged replisomes~\cite{RPD+08}.  In the
approximation of constant replication fork velocity along the mother
DNA, replication time in this experiment and genomic arclength
distance have equal scaling. Hence, neglecting the global movements due
to chromosomal segregation, the replisome's anomalous diffusion
exponent $\alpha_R$ measures the effective fractal dimension $D_{f,R}$
of the replicating DNA. Specifically, the scaling $MSD(t) \sim
t^{\alpha_R}$ implies $\alpha_R \sim 2/D_{f,R}$.
Obviously, this measured $D_{f,R}$ in principle contains errors due to
fluctuations of a stationary background as well as large scale effects
associated with coherent restructuring of the nucleoid.  The latter
processes are relevant for segregation, but in the initial phases of
replication one can assume that they can be disregarded, as we are in
the presence of a ``weak perturbation'' of the stationary
(non-replicating) structure.  In such a case, the time fluctuations of
the fork velocity and the steady state anomalous diffusion of genetic
loci would be the main corrections to be taken into account in order
to estimate $D_{f}$ from the measured $D_{f,R}$.  Quite interestingly
the experimental estimates of the exponent $\alpha_{R}$
from~\cite{RPD+08} are $\alpha_{R} \sim 0.66$ and $0.58$ for
experiments with $30$s and $5$min time lapses respectively.  If
these numbers were confirmed by further measurements and accurate data
analysis, they would support the hypothesis that $D_{f} \sim 3$ for a
range of chromosomal scales, independently on the reasoning presented
above, based on the DeGennes scaling relation $z = 2 + D_{f}$.

Clearly, for the nucleoid one expects a structure with a range of
fractal dimensions in different scale regimes, as suggested by the
case of chromatin and the ribosome. The existence of a range of
fractal dimensions is also supported by the microscopy results for
fluorescent pairs of loci reporting a linear correlation between loci
distance in the cell and along the chromosome~\cite{Wiggins2010}.
The linear correlation could result from any sequentially ordered
segregation process generating a uniform mass density.  For example,
one might consider a periodic winding of the
chromosome~\cite{kepes1,Mathelier2010}, that could be produced by a
segregation choreography of the type observed in \emph{B.~subtilis} as
well as in \emph{E.~coli}. However, this needs to be reconciled with
the observed subdiffusion.
As for the case of chromatin, a hierarchical structure is very
logical, since some chromosomal functions, such as transcription,
replication and DNA repair, require a certain degree of plasticity,
and are not compatible with full compaction at all scales and at all
times.
Thus, the linear correlation of loci subcellular position and genomic
distance can be consistent with $D_{f} \approx 3$, but further work is
needed to determine the range of spatial scales where these properties
apply.

The main objection against the fractal-globule as a long-lasting
transient state supported by topological constraints is the ubiquitous
presence of topoisomerases, DNA enzymes able to cut and paste strands
and thus easily resolve these constraints. An interesting theoretical
study has focused on the entanglement of  tethered
rings~\cite{Marko2009}; it is argued that entanglements would
``condense'', i.e. aggregate in space, in physically relevant
situations, which, in presence of enzymes, would facilitate further
the resolution of topological constraints.
This kind of objection holds for both the eukaryotic and the
prokaryotic case.
%
It cannot be excluded that the non-equilibrium constraints leading to
the fractal structure observed in Hi-C experiments are caused by
something else, or more in general, and more plausibly, that a
different physical process than simple topological constraints leads
to the observed phenomenology. However, the generic reasoning
presented above for connecting $D_f$ and $\alpha$ might be robust with
respect to these considerations.

Other approaches to chromatin organization aim at reproducing the
interlocus distance $R(s)$ and the distribution of interacting loci
$P_c(s)$ with alternative polymer models,
such as a collapsing self-avoiding walk in a solution of organizing
proteins which can bind and act as discrete self-attraction points,
representing organizing proteins~\cite{Nicodemi2009}. The spirit of
this kind of study is to go beyond a dominant role of entropy, and
take more seriously the ``energetic part'' of the free energy, and in
particular the organizing proteins.  The consequences of this
hypothesis are explored by analyzing the resulting equilibrium
structures for the polymer.
Nicodemi and coworkers have recently found that such a model polymer
could be close the $\theta$ point for empirically relevant protein
concentrations, and small variation of the concentration of binding
proteins around this state could recapitulate a considerable part of
the observed phenomenology of nuclear eukaryotic
DNA~\cite{Nicodemi2011}.

Finally, as mentioned above, it is also worthwhile to consider whether
equilibrium statistical mechanics is even the proper starting point to
understand the structure of the bacteria nucleoid. Cells expend
a considerable amount of energy maintaining steady-state,
non-equilibrium environments. A classic example is the membrane
potential, whose existence requires an elaborate mechanism for pumping
protons through the membrane. Given that the genomic information is
arguably the most important part of the cell, containing both the
instructions for the current cell and the inheritable information for
the next generation, it seems unlikely that bacteria have evolved such
that the structure of the nucleoid is resigned to equilibrium.

\section{Conclusions. }

To conclude, we briefly review some of the main features of the
partial and emerging picture of the nucleoid, from the physics
viewpoint.
All these problems still need to be understood in a quantitative
framework for bacterial physiology~\cite{Scott2011}, and in particular
for varying growth rates (and subcellular compositions) and during
adaptation to different growth conditions~\cite{Muskhelishvili2010}.

A first problem that can be isolated is the explanation of its
compaction/condensation properties.
Likely mechanisms that can influence (positively or negatively)
nucleoid condensation (and more than one can be at play) include (i)
supercoiling, bending and looping, in interplay with binding of NAPs,
which can cause punctual or polymer-like links building aggregation
foci (H-NS, MatP) and stabilize a ramified plectonemic loop
structure (Fis), (ii) consequences of molecular crowding, in the form
of both phase separation and depletion interactions, and (iii)
(nonequilibrium) segregation after replication, which could be induced
by different physical processes.

A second, related problem is the \emph{geometry} of the nucleoid,
which requires  understanding how the subunits are
arranged in the cell at different scales and times. It is likely that the
organization principles in a given cell state (determined e.g. by
growth rate and growth phase) are different at different scales.
At the micron scale, the experiments seem to converge towards a
linearly-arranged sausage-shaped structure, sometimes wrapped by the
Ter region, and the main outstanding questions seem to relate to the
physical mechanisms behind segregation and its choreography.
Below this scale, the existence of macrodomains and transcription foci
still elude a physical explanation, which could be microphase
separation stabilized by short-ranged attractions of chemical
(organizing proteins such as MatP) or of entropic (ribosome-induced
depletion interactions) origin.
At an even smaller scale, an organization in blobs or fibers seems to
be equally elusive, despite the existence of numerous pieces of evidence
for different aspects of NAP binding and plectonemic loop formation
and stabilization.

Finally, it is important to point out that within the layered
information given here there lies more than one unresolved question.
For example, if macrodomains are microphases structured by protein
binding, then certainly these proteins must play an important role in
the configurational entropy of the nucleoid, which is not considered
in the arguments concerning entropy-driven chromosome
segregation. Also, if the genome is compacted (at least in a range of
scales) in a fractal or conventional globule configuration by
attractive interactions of entropic or energetic origin, this will
greatly affect its entropy, and thus its mechanical properties, loci
dynamics and the interactions between segregating chromosomes.
Equally important, the supercoiling-independent size of structural units
measured for purified nucleoids (whose size varies with supercoiling)
appears challenging for theoretical explanations.
While we are certainly far from a coherent and consistent physical
description of the nucleoid, there is a clear abundance of existing
data and many ongoing experiments merging quantitative biophysics and
high-throughput molecular biology. These emerging results, together
with the fragmented but partially successful modeling approaches, make
us believe that we might be on the verge of resolving at least some of
the existing issues regarding the physics of the bacterial nucleoid.

\begin{acknowledgments}
  We are very grateful to Bianca Sclavi for discussions, feedback, and
  help with the revision of this manuscript, and to Mario Nicodemi,
  Andrea Parmeggiani, Eric Siggia, Georgi Muskhelishvili, Ivan Junier,
  Zhicheng Long, Avelino Javer, Matteo Osella, Matthew Grant, and
  Eileen Nugent for useful discussions. We also thank Christine Hardy
  for kindly allowing us to reprint Fig.~3A. This work was supported by
  the International Human Frontier Science Program Organization, grant
  RGY0069/2009-C.
\end{acknowledgments}

\bibliography{bibs}

\end{document}